\documentclass[a4paper,fleqn,usenatbib]{mnras}

\usepackage[T1]{fontenc}
\usepackage{ae,aecompl}

\usepackage{graphicx}	
\usepackage{amsmath}	
\usepackage{amssymb}	

\renewcommand{\theenumi}{\arabic{enumi})}

\title[Dynamical Vertical Structure of TDE Discs]{Dynamical structure of highly eccentric discs with applications to tidal disruption events}

\author[E. M. Lynch and G. I. Ogilvie]{
Elliot M. Lynch \thanks{E-mail: eml52@cam.ac.uk} and Gordon I. Ogilvie
\\
Department of Applied Mathematics and Theoretical Physics, University of Cambridge, Centre for Mathematical Sciences, \\ Wilberforce Road, Cambridge CB3 0WA, UK\\
}

\date{Accepted version}

\pubyear{2019}

\begin{document}
\label{firstpage}
\pagerange{\pageref{firstpage}--\pageref{lastpage}}
\maketitle

\begin{abstract}
Whether tidal disruption events circularise or accrete directly as highly eccentric discs is the subject of current research and appears to depend sensitively on the disc thermodynamics. One aspect of this problem that has not received much attention is that a highly eccentric disc must have a strong, non-hydrostatic variation of the disc scale height around each orbit. As a complement to numerical simulations carried out by other groups, we investigate the dynamical structure of TDE discs using the nonlinear theory of eccentric accretion discs. In particular, we study the variation of physical quantities around each elliptical orbit, taking into account the dynamical vertical structure, as well as viscous dissipation and radiative cooling. The solutions include a structure similar to the nozzle-like structure seen in simulations. We find evidence for the existence of the thermal instability in highly eccentric discs dominated by radiation pressure.  For thermally stable solutions many of our models indicate a failure of the $\alpha-$prescription for turbulent stresses. We discuss the consequences of our results for the structure of eccentric TDE discs. 

\end{abstract}




\section{Introduction}

Tidal disruption events (TDEs) are transient phenomena where an object on a nearly parabolic orbit passes within the tidal radius and is disrupted by the tidal forces, typically a star being disrupted by a supermassive black hole (SMBH). More recently TDEs have been discussed in the context of disruption of a planetesimal by a white dwarf \citep{Manser16,Cauley18,Miranda18,Malamud19b,Malamud19a}. The material from the disrupted object is emplaced onto a range of orbits some of which will be bound to the central object. The bound material forms a highly eccentric disc that accretes onto the central object. In the classic TDE model of \citet{Rees88} mass is uniformly distributed in orbital energy. In this model, bound material undergoes orbital intersection, resulting in shocks that rapidly circularise the material as it returns to pericentre. The energy lost by the material during circularisation is radiated, resulting in a luminosity with a characteristic power-law decay with exponent $-5/3$ \citep{Rees88} (corrected in \citet{Phinney89}).

The possibility that TDEs do not circularise but accrete more directly onto the black hole has been put forward to explain some discrepancies between observations and theory. It has been suggested that the direct accretion scenario helps explain the lower observed energies \citep{Svirski17} and shallower late-time luminosity decay laws \citep{Auchettl17} compared to the classic \citet{Rees88} model. Separately \citet{Liu17} and \citet{Cao18} proposed modelling optical TDE spectra as a reprocessing of X-rays in a eccentric disc with $e\approx0.97$, in order to fit the double peaked structure of the H$\alpha$ line. \citet{Holoien19} also showed that the H $\alpha$ emission of PS18kh indicated non-axisymmetric structures in the TDE disc, consistent with either an eccentric disc or a spiral arm. However recently \citet{Hung20} have claimed a much lower eccentricity of $e \approx 0.1$ for the TDE AT 2018hyz.

The hydrodynamic simulations of \citet{Hayasaki16} and \citet{Bonnerot16} have shown that, when apsidal precession is weak, circularisation of TDE discs depends on the assumed thermodynamics, in particular how efficient the disc is at radiating energy generated by shocks or turbulence. This contrasts with the rapid circularisation which can occur when apsidal precession is strong \citep{Hayasaki13,Sadowski16}. In their radiatively efficient case the disc loses any energy generated in shocks or from viscous heating (the behaviour of the fluid being polytropic in \citet{Hayasaki16} and isothermal in \citet{Bonnerot16}). In this limit circularisation is not efficient and the disc forms a thin eccentric ring. In the opposite radiatively inefficient limit the disc satisfies the appropriate thermal energy equation so that kinetic energy lost in shocks and to viscosity is converted to thermal energy in the disc. In this limit the disc rapidly circularises and heats up to form a thick accretion torus \citep{Hayasaki16,Bonnerot16}. Simulations by \citet{Shiokawa15}, also radiatively inefficient, additionally allowed for a temperature-dependent first adiabatic exponene to account for contributions to the pressure from both the radiation and the gas. The disc formed in \citet{Shiokawa15} is thick as in the other radiatively inefficient models; however, the disc does not rapidly circularise and maintains moderate eccentricities throughout the disc. To date simulations of circularisation with explicit cooling physics have not been carried out. 

TDE discs are expected to be dominated by radiation pressure (e.g. see calculations in \citet{Loeb97,Shen14}), and the thermal stability of a radiation pressure dominated disc is currently an open question. For a circular disc with an alpha model \citep{Shakura73} of the viscous stress a radiation pressure dominated disc is thermally unstable when the stress scales with the total (radiation + gas) pressure \citep{Shakura76,Pringle76,Piran78}, but thermally stable when stress scales with gas pressure \citep{Meier79,Sakimoto81}. The $\alpha-$viscosity prescription is a model of the turbulent stress which arises from the action of the magnetorotational instability (MRI, \citet{Balbus91,Hawley91}). While early MRI simulations including thermal physics tended to support thermal stability \citep{Hirose09}, more recent calculations \citep{Latter12,Jiang13, Mishra16,Ross17}, where the box size is large enough to contain the most unstable modes, support stress scaling with total pressure and imply thermal instability in radiation pressure dominated discs. The thermal instability in a MRI turbulent circular disc results in either thermal runaway \citep{Latter12,Jiang13,Ross17} or collapse to a thin gas pressure dominated disc \citep{Jiang13,Mishra16}.

Eccentric discs, in which the dominant motion consists of elliptical Keplerian orbits, occur in a wide variety of astrophysical situations. Discs with extreme eccentricities approaching $1$ have been discussed in the context of TDEs \citep{Guillochon14,Piran15,Krolik16,Svirski17}. Very little attention has been paid to the vertical structure of such extreme discs, which is more complicated than that of circular discs and is likely to be important in determining how these discs evolve, as well as their observational properties.

Unlike circular discs, the vertical structure of an eccentric disc cannot be decoupled from the horizontal fluid motion. Eccentric discs cannot maintain hydrostatic equilibrium owing to the variation of vertical gravity around an elliptical orbit. This drives vertical motions in the disc, with the disc scale height varying around the orbit; this variation modifies the pressure and affects the horizontal dynamics \citep{Ogilvie01,Ogilvie08,Ogilvie14}.

In this paper we apply eccentric disc theory to the highly eccentric TDE discs in order to determine the dynamics of their vertical structure. The vertical motion, radiative physics and viscous heating are important for understanding dissipative processes in TDEs, which is key to determining whether TDEs circularise.

This paper is structured as follows. In Section \ref{orbital coords} we discuss the geometry of eccentric discs and restate the coordinate system of \citet{Ogilvie19}. In Section \ref{breathinmode} we present the equations for the dynamical vertical structure in this coordinate system and in Section \ref{sol method} we discuss how we go about solving these equations. In Sections \ref{tde sols} and \ref{radgas} we apply our model to the highly eccentric discs expected in TDEs. In Section \ref{nozzle} we focus on the pericentre passage and compare our results to the nozzle structures found in simulations.  In Section \ref{discuss} we discuss complications that arise when using a $\gamma=4/3$ perfect gas to model a radiation pressure dominated disc (Section \ref{rad gas perf gas comp}), the thermal stability of our solutions (\ref{stabil}), the breakdown of the $\alpha-$prescription (\ref{alpha presecription issues}) and possible implications of our model for existing work (\ref{reflect imp}). We present our conclusions in Section \ref{conc} and additional mathematical details in the appendices.

\section{Orbital Coordinates} \label{orbital coords}

In this paper we assume that the dominant motion in a TDE disc consists of elliptical Keplerian orbits, subject to relatively weak perturbations from relativistic precessional effects, pressure and viscosity. This assumption should be valid if disruption of the star occurs sufficiently far from the black hole, but will be poor for TDEs with a strong departure from Newtonian gravity.

Let $(r,\phi)$ be polar coordinates in the disc plane. The polar equation for an elliptical Keplerian orbit of semimajor axis $a$, eccentricity $e$ and longitude of periapsis $\varpi$ is

\begin{equation}
r = \frac{a (1 - e^2)}{1 + e \cos f} \quad ,
\end{equation} 
where $f = \phi - \varpi$ is the true anomaly. A planar eccentric disc involves a continuous set of nested elliptical orbits. The shape of the disc can be described by considering $e$ and $\varpi$ to be functions of $a$. The derivatives of these functions are written as $e_a$ and $\varpi_a$, which can be thought of as the eccentricity gradient and the twist, respectively. The disc evolution is then described by the slow variation in time of the orbital elements $e$ and $\varpi$ due to secular forces such as pressure gradients in the disc and departures from the gravitational field of a Newtonian point mass.

In this work we adopt the (semimajor axis $a$, eccentric anomaly $E$) orbital coordinate system described in \citet{Ogilvie19}. The eccentric anomaly is related to the true anomaly by

\begin{equation}
\cos f = \frac{\cos E - e}{1 - e \cos E} , \quad  \sin f = \frac{\sqrt{1 - e^2} \sin E}{1 - e \cos E}
\end{equation}
and the radius can be written as

\begin{equation}
r = a (1 - e \cos E) \quad .
\end{equation}

The area element in the orbital plane is given by $d A = (a n/2) J \, d a \, d E$ where $J$ is given by

\begin{equation}
J = \frac{2}{n} \left[\frac{1 -e (e + a e_a)}{\sqrt{1 - e^2}} - \frac{a e_a}{\sqrt{1 - e^2}} \cos E - a e \varpi_a \sin E\right] ,
\end{equation} 
which  corresponds to the Jacobian of the $(\Lambda,\lambda)$ coordinate system of \citet{Ogilvie19}. Here $n = \sqrt{\frac{G M_{\bullet}}{a^3}}$ is the mean motion with $M_{\bullet}$ the mass of the black hole. The Jacobian can be written in terms of the orbital intersection parameter $q$ of \citet{Ogilvie19}:

\begin{equation}
J = \frac{2}{n} \frac{1 -e (e + a e_a)}{\sqrt{1 - e^2}}  (1 - q \cos(E - E_0)) \quad ,
\end{equation}
where $q$ is given by

\begin{equation}
q^2 = \frac{(a e_a)^2 + (1 - e^2) (a e \varpi_a)^2}{[1 - e (e + a e_a)]^2} \quad ,
\end{equation}
and we require $|q| < 1$ to avoid an orbital intersection \citep{Ogilvie19}. The angle $E_0$, which determines the location of maximum horizontal compression around the orbit, is determined by the relative contributions of the eccentricity gradient and twist to $q$:

\begin{equation}
\frac{a e_a}{1 - e(e + a e_a)} = q \cos E_0 \quad. 
\end{equation}

Additionally it can be useful to rewrite time derivatives, following the orbital motion, in terms of the eccentric anomaly:

\begin{equation}
\frac{\partial}{\partial t} = \frac{n}{(1 - e \cos E)} \frac{\partial}{\partial E} \quad .
\end{equation}

\section{Derivation of the equations of vertical structure, including thermal physics} \label{breathinmode}

A local model of a thin, Keplerian eccentric disc was developed in \citet{Ogilvie14}. For a disc that includes viscosity and radiative cooling a local model can be obtained in a similar manner. We use the $(a,E)$ orbital coordinates of \citet{Ogilvie19} as described above rather than those in \citet{Ogilvie14}. The equations, formulated in a frame of reference that follows the elliptical orbital motion, are the vertical equation of motion,

\begin{equation}
\frac{D v_z}{D t} = - \frac{G M_{\bullet} z}{r^3} - \frac{1}{\rho} \frac{\partial}{\partial z} \left(p - T_{z z} \right),
\label{vertical laminar flow}
\end{equation}
the continuity equation,

\begin{equation}
\frac{D \rho}{D t} = - \rho \left( \Delta + \frac{\partial v_z}{\partial z} \right) ,
\end{equation}
and the thermal energy equation, 

\begin{equation}
\frac{D p}{D t} =  - \Gamma_1 p \left( \Delta + \frac{\partial v_z}{\partial z} \right) + (\Gamma_3 - 1) \left( \mathcal{H} - \frac{\partial F}{\partial z} \right) ,
\end{equation}
where, for horizontally invariant ``laminar flows'',

\begin{equation}
\frac{D}{D t} = \frac{\partial}{\partial t} + v_z \frac{\partial}{\partial z}
\end{equation}
is the Lagrangian time-derivative,

\begin{equation}
\Delta = \frac{1}{J} \frac{dJ}{dt}
\end{equation}
is the divergence of the orbital velocity field, which is a known function of $E$ that depends on $e$, $q$ and $E_0$. $F = F_{\rm rad} + F_{\rm ext}$ is the total vertical heat flux with

\begin{equation}
F_{\rm rad} = - \frac{16 \sigma T^3}{3 \kappa \rho} \frac{\partial T}{\partial z} 
\end{equation}
being the vertical radiative heat flux and $F_{\rm ext}$ containing any additional contributions to the heat flux (such as from convection or turbulent heat transport). 

\begin{equation}
T_{z z} =2 \mu_s \frac{\partial v_z}{\partial z} + \left(\mu_b - \frac{2}{3} \mu_s \right) \left(\Delta + \frac{\partial v_z}{\partial z} \right)
\end{equation}
is the $z z$ component of the viscous stress tensor and $\mathcal{H}$ is the viscous heating rate per unit volume. $\mu_s$ and $\mu_b$ are the dynamical shear and bulk viscosities. $\Gamma_1$ and $\Gamma_3$ are the first and third adiabatic exponents.

We assume the pressure in the disc, which includes contributions from radiation and a perfect gas with a ratio of specific heats $\gamma$, is given by

\begin{equation}
p = p_{r} + p_{g} = \frac{4 \sigma}{3 c} T^4 + \frac{\mathcal{R} \rho T}{\mu} \quad .
\end{equation}
We define $\beta_r$ to be the ratio of radiation to gas pressure:

\begin{equation}
\beta_{r} := \frac{p_r}{p_g} =   \frac{4 \sigma \mu}{3 c \mathcal{R}}  \frac{T^3}{\rho} \quad .
\label{betar def}
\end{equation}
We assume a constant opacity $\kappa$, applicable to the electron-scattering opacity expected in a TDE. 

We further assume an alpha viscosity, 

\begin{equation}
\mu_{s,b} = \alpha_{s,b} \frac{p_v}{\omega_{\rm orb}} ,
\end{equation}
where $\alpha_{s,b}$ are dimensionless coefficients and $\omega_{\rm orb}$ is some characteristic frequency of the orbital motion. Although we include it for completeness, all the models considered in this paper have $\alpha_b = 0$. In the $\alpha-$viscosity prescription the stress scales with some pressure $p_v$; for the radiation-gas mixture it is not clear whether this pressure should be the total pressure, the gas pressure or some average of the two. For the pure gas there is only one pressure so $p_v = p$.

It is very unclear what the correct frequency to use should be. For instance we could have $\omega_{\rm orb} = n = (G M_{\bullet}/a^3)^{1/2}$, the mean motion, $\omega_{\rm orb} = \Omega = n \sqrt{1 - e^2} (1 - e \cos E)^{-2}$, the angular frequency or $\omega_{\rm orb} = n (1 - e \cos E)^{-3/2}$, the circular frequency at that radius. In fact \citet{Ogilvie01} used a different scaling with $\omega_{\rm orb} = n (1 - e^2)^{-3/2}$. In a circular disc this distinction is unimportant as $n = \Omega = \Omega_c$, but for a highly eccentric disc $\Omega, \Omega_c \gg n$ at pericentre, so this makes a significant difference to the strength of the viscosity. In this paper we choose $\omega_{\rm orb} = n$ simply because this is independent of the eccentricity and eccentric anomaly, this choice is equivalent to \citet{Ogilvie01} with an appropriately rescaled $\alpha$. In practice we can obtain the same qualitative behaviours for different choices of $\omega_{\rm orb}$, but with slightly different disc parameters. Given the many uncertainties surrounding the applicability of the $\alpha$-prescription, and the behaviour of MRI turbulence, in highly eccentric discs, determining which $\omega_{\rm orb}$ most closely mimics MRI is beyond the scope of this paper.

The viscous heating rate is 

\begin{equation}
\mathcal{H} = f_{\mathcal{H}} (t) n p_v ,
\end{equation}
where $f_{\mathcal{H}}$ is a complicated dimensionless expression given in Appendix \ref{heat func expl}.

We consider two situations, a pure gas disc with $F_{\rm ext} = 0$ and $\Gamma_{1} = \Gamma_{3} = \gamma$ and a radiation+gas mixture where $F_{\rm ext}$ is assumed to be from convective or turbulent heat transport and the first and third adiabatic exponents are given by \citep{Chandrasekhar67}

\begin{equation}
\Gamma_1 = \frac{1 + 12 (\gamma - 1) \beta_r + (\gamma - 1) (1 + 4 \beta_r)^2 }{(1 + \beta_r) (1 + 12 (\gamma - 1) \beta_r)} \quad ,
\label{Gamma 1 radmix}
\end{equation}

\begin{equation}
\Gamma_3 = 1 + \frac{(\gamma - 1) (1 + 4 \beta_r)}{1 + 12 (\gamma - 1) \beta_r} \quad .
\label{Gamma 3 def}
\end{equation}

We propose a separable solution of the form

\begin{align}
\begin{split}
\rho &= \hat{\rho} (t) \tilde{\rho} (\tilde{z}) , \\
p &= \hat{p} (t) \tilde{p} (\tilde{z}) , \\
T &= \hat{T} (t) \tilde{T} (\tilde{z}) , \\
F &= \hat{F} (t) \tilde{F} (\tilde{z}) , \\
v_{z} &= \frac{d H}{d t}\tilde{z} , \\
\end{split}
\end{align}
where 

\begin{equation}
\tilde{z} = \frac{z}{H(t)}
\end{equation}
is a Lagrangian variable that follows the vertical expansion of the disc, i.e. $D\tilde{z}/Dt = 0$, and the quantities with the tildes satisfy the dimensionless equations of vertical structure, 

\begin{equation}
\frac{d \tilde{p}}{d \tilde{z}} = - \tilde{\rho} \tilde{z} ,
\label{pressure vert strut}
\end{equation}

\begin{equation}
\frac{d \tilde{F}}{d \tilde{z}} = \lambda \tilde{p} ,
\label{flux vert struct}
\end{equation}

\begin{equation}
\frac{d \tilde{T}}{d \tilde{z}} = -\tilde{\rho} \tilde{T}^{- 3} \tilde{F}_{\rm rad} ,
\label{temp vert strut}
\end{equation}

\begin{equation}
\tilde{p} = \tilde{\rho} \tilde{T} .
\label{eos vert strut}
\end{equation}
Here $\lambda$ is a dimensionless eigenvalue of the vertical structure problem when appropriate boundary conditions are imposed and corresponds to a dimensionless cooling rate. As the only influence these equations have on what follows is through $\lambda$ we leave a discussion of their solutions and implications to TDE structure to Appendix \ref{general vert struct}.

The separated solution works provided that

\begin{align}
\begin{split}
\frac{d^2 H}{d t^2} &= - \frac{G M_{\bullet}}{r^3} H + \frac{\hat{p}}{\hat{\rho} H} \\
&\times \left[ 1 - \frac{2 \alpha_s}{n H} \frac{\hat{p}_v}{\hat{p}} \frac{d H}{d t} - \frac{(\alpha_b - \frac{2}{3} \alpha_s)}{n} \frac{p_v}{p} \left(\Delta + \frac{1}{H} \frac{d H}{d t}\right)\right] ,
\end{split}
\end{align}

\begin{equation}
\frac{d \hat{\rho}}{d t} = - \hat{\rho} \left ( \Delta + \frac{1}{H} \frac{d H}{d t} \right) ,
\end{equation}

\begin{equation}
\frac{d \hat{p}}{d t} = - \Gamma_1 \hat{p} \left ( \Delta + \frac{1}{H} \frac{d H}{d t} \right) + (\Gamma_3 - 1)\left(  f_{\mathcal{H}} n \hat{p}_v - \lambda \frac{\hat{F}}{H} \right) ,
\end{equation}

\begin{equation}
\hat{F} = \frac{16 \sigma \hat{T}^{4}}{3 \kappa \hat{\rho} H} \quad ,
\end{equation}

\begin{equation}
\hat{p} = (1 + \beta_r) \frac{\mathcal{R} \hat{\rho} \hat{T}}{\mu} \quad ,
\end{equation}
where $\beta_r = 0$ for the pure gas and 

\begin{equation}
\beta_r = \frac{4 \sigma \mu}{3 c \mathcal{R}}  \frac{\hat{T}^3}{\hat{\rho}}
\end{equation}
for the radiation-gas mixture.

Note that the surface density and vertically integrated pressure are

\begin{equation}
\Sigma = \hat{\rho} H , \quad P = \hat{p} H , \quad  P_v = \hat{p}_v H .
\end{equation}
The vertically integrated heating and cooling rates are

\begin{equation}
f_{\mathcal{H}} n P_v ,\quad \lambda \hat{F} \quad .
\end{equation}
The cooling rate can also be written as

\begin{equation}
\lambda \hat{F} = 2 \sigma \hat{T}_s^4
\end{equation}
where $\hat{T}_{s} (t)$ is a representative surface temperature defined by

\begin{equation}
\hat{T}^{4}_s = \frac{8 \lambda}{3} \frac{\hat{T}^4}{\tau}
\end{equation}
and

\begin{equation}
\tau = \kappa \Sigma
\end{equation}
is the optical thickness.

We then have

\begin{align}
\begin{split}
\frac{1}{H} \frac{d^2 H}{d t^2} &= - \frac{G M_{\bullet}}{r^3} + \frac{P}{\Sigma H^2} \\
&\times \Biggl[1 - \frac{2 \alpha_s}{n H} \frac{P_v}{P} \frac{d H}{d t} - \frac{(\alpha_b - \frac{2}{3} \alpha_s)}{n} \frac{P_v}{P} \left (\Delta + \frac{1}{H} \frac{d H}{d t} \right) \Biggr] ,
\end{split}
\end{align}

\begin{equation}
J \Sigma = \mathrm{constant},
\end{equation}

\begin{equation}
 \frac{d P}{d t} = - \Gamma_1 P\left( \Delta + \frac{1}{H} \frac{d H}{d t} \right) + \frac{P}{H} \frac{d H}{d t}  + (\Gamma_3 - 1 ) n P \left( f_{\mathcal{H}} \frac{P_v}{P}  - \frac{\lambda \hat{F}}{n P} \right) ,
\end{equation}
with

\begin{equation}
\frac{\lambda \hat{F}}{n P} = \lambda \frac{16 \sigma (\mu/\mathcal{R})^{4}}{3 \kappa n} P^{3} \Sigma^{- 5} (1 + \beta_r)^{-4} \quad .
\end{equation}

For a standard circular disc we have $f_{\mathcal{H}} = \frac{9}{4} \alpha_s$ and hydrostatic balance, $P/\Sigma H^2 = G M_{\bullet}/r^3 = n^2$. These determine the equilibrium values, $H^{\circ}$ and $\hat{T}^{\circ}$, etc., for a reference circular orbit, with the same mass and semimajor axis, in terms of $\Sigma$ and $n$. When $p_v=p$ for the radiation-gas mixture, there are two physical solution branches; for what follows the reference state is the gas pressure dominated branch.

Scaling $H$ by $H^{\circ}$, $\hat{T}$ by $T^{\circ}$, $t$ by $1/n$, $r$ by $a$ and $J$ by $2/n$ we obtain the dimensionless version

\begin{align}
\begin{split}
\frac{\ddot{H}}{H} &= -(1 - e \cos E)^{-3} + \frac{T}{H^2} \frac{1 + \beta_r}{1 + \beta_{r}^{\circ} } \\
&\times \Biggl[ 1 - 2 \alpha_s \frac{P_v}{P} \frac{\dot{H}}{H} - (\alpha_b - \frac{2}{3} \alpha_s) \frac{P_v}{P} \left(\frac{\dot{J}}{J} + \frac{\dot{H}}{H}\right)\Biggr] ,
\end{split}
\label{scale height equation}
\end{align}

\begin{align}
\begin{split}
 \dot{T} &= - (\Gamma_3 - 1) T \left(\frac{\dot{J}}{J} + \frac{\dot{H}}{H} \right) \\
&+ (\Gamma_3 - 1) \frac{1 + \beta_r}{1 + 4 \beta_r} T \left( f_{\mathcal{H}} \frac{P_v}{P} - \frac{9}{4} \alpha_s \frac{P^{\circ}_v}{P^{\circ}} \frac{1 + \beta_{r}^{\circ} }{1 + \beta_r} J^2 T^{3} \right) ,
\end{split}
\label{thermal energy equation T}
\end{align} 
where a dot over a letter indicates a derivative with respect to the rescaled time. We have written the thermal energy equation in terms of the temperature. The factor $\frac{\Gamma_{3} - 1}{1 + 4 \beta_r} \propto \frac{1}{c_{V}}$ where $c_{V}$ is the specific heat capacity at constant volume. $\beta_r$ can be obtained through

\begin{equation}
\beta_r = \beta_{r}^{\circ} J H T^3 \quad ,
\end{equation}
where we have introduced $\beta_{r}^{\circ} $, which is the $\beta_{r}$ of the reference circular disc. The equilibrium values of the reference circular disc $H^{\circ}$, $\hat{T}^{\circ}$, etc., are determined by $\beta_r^{\circ}$ and $n$.

We introduce the (nondimensionalised) specific entropy

\begin{equation}
s := \ln ( J H T^{1/(\gamma - 1)} ) + 4 \beta_r  \quad ,
\label{kentrop radmix}
\end{equation}
which includes contributions from the gas and radiation. This evolves according to

\begin{equation}
\dot{s} = (1 + \beta_r) \left( f_{\mathcal{H}} \frac{P_v}{P} - \frac{9}{4} \alpha_s \frac{P_v^{\circ}}{P^{\circ}} \frac{1 + \beta_{r}^{\circ} }{1 + \beta_r} J^2 T^3 \right)
\label{thermal energy equation}
\end{equation}
which we solve for the pure gas case instead of Equation \ref{thermal energy equation T}. For the radiation-gas mixture we solve Equation \ref{thermal energy equation T} to avoid having to invert Equation \ref{kentrop radmix} for the temperature at each timestep.

\section{Periodic Attractor Solution to the Vertical Structure Equations} \label{sol method}

In our numerical setup we solve Equations \ref{scale height equation} and \ref{thermal energy equation T} for the radiation-gas mixture and Equations \ref{scale height equation} and \ref{thermal energy equation} for a pure gas where $\Gamma_1 = \Gamma_3 =  \gamma$.

\begin{figure}
\includegraphics[trim=10 0 10 10,clip,width=\linewidth]{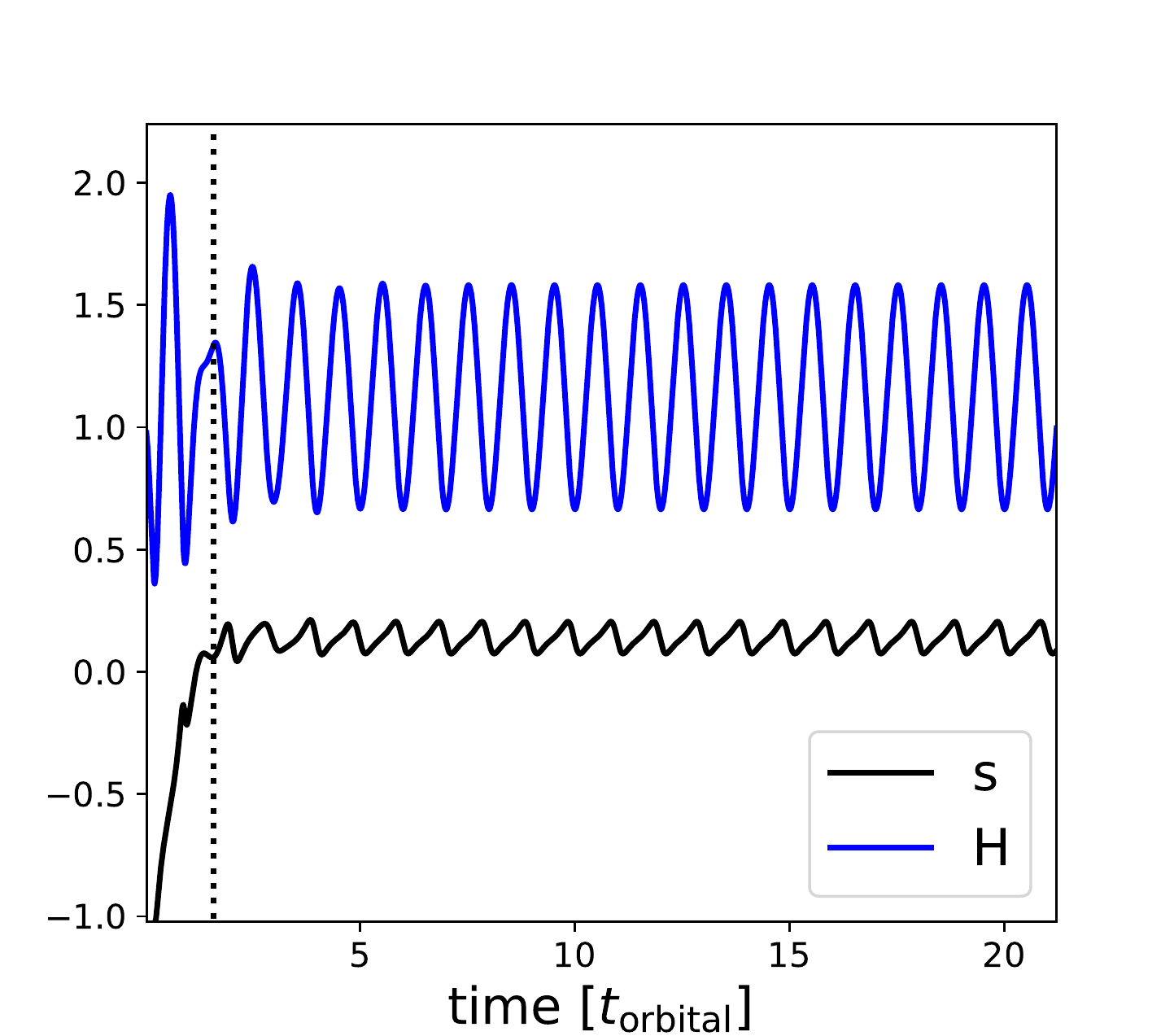}
\caption{Convergence of the scaleheight and entropy towards a periodic solution. The vertical dashed line is at one thermal timescale after the start. The parameters of this pure gas model are $\gamma = 5/3$, $\alpha_s = 0.1$, $\alpha_b = 0.0$, $e=0.3$, $q=0.3$,  $E_0=0$.} 
\label{example convergence}
\end{figure}

We integrate these equations forward in eccentric anomaly using a 4th order Runge-Kutta method with a control on the relative error on $H$. We expect a $2 \pi$-periodic solution to act as an attractor or limit cycle for trajectories in phase space so that from our initial guess ($T=H=1$, $\dot{H}=0$ corresponding to the hydrostatic solution in the reference circular disc) the solution should relax to this after sufficient time. Figure \ref{example convergence} shows the relaxation towards a periodic solution of one such disc.

The $2 \pi$-periodic solution generally involves a nonlinear forced breathing mode which is maintained against dissipation by periodic forcing by the variation of $J$ (for a disc with $q \ne 0$) and the vertical gravity around the orbit. It corresponds to a solution which is steady but non-axisymmetric in an inertial frame of reference. Additional free oscillations of the breathing mode, which have been studied in circular discs \citep{Lubow81,Ogilvie02b}, can also exist in an eccentric disc; however, without an excitation mechanism we expect these to damp given sufficient time. In fact we find that if numerical errors are not controlled carefully these can excite the free mode so that instead of a periodic attractor there is a quasiperiodic attractor solution, being a superposition of the free and forced modes. If the numerical errors are controlled sufficiently the free mode will damp until it reaches the amplitude of the numerical errors.

\begin{figure}
\includegraphics[trim=0 0 10 10,clip,width=\linewidth]{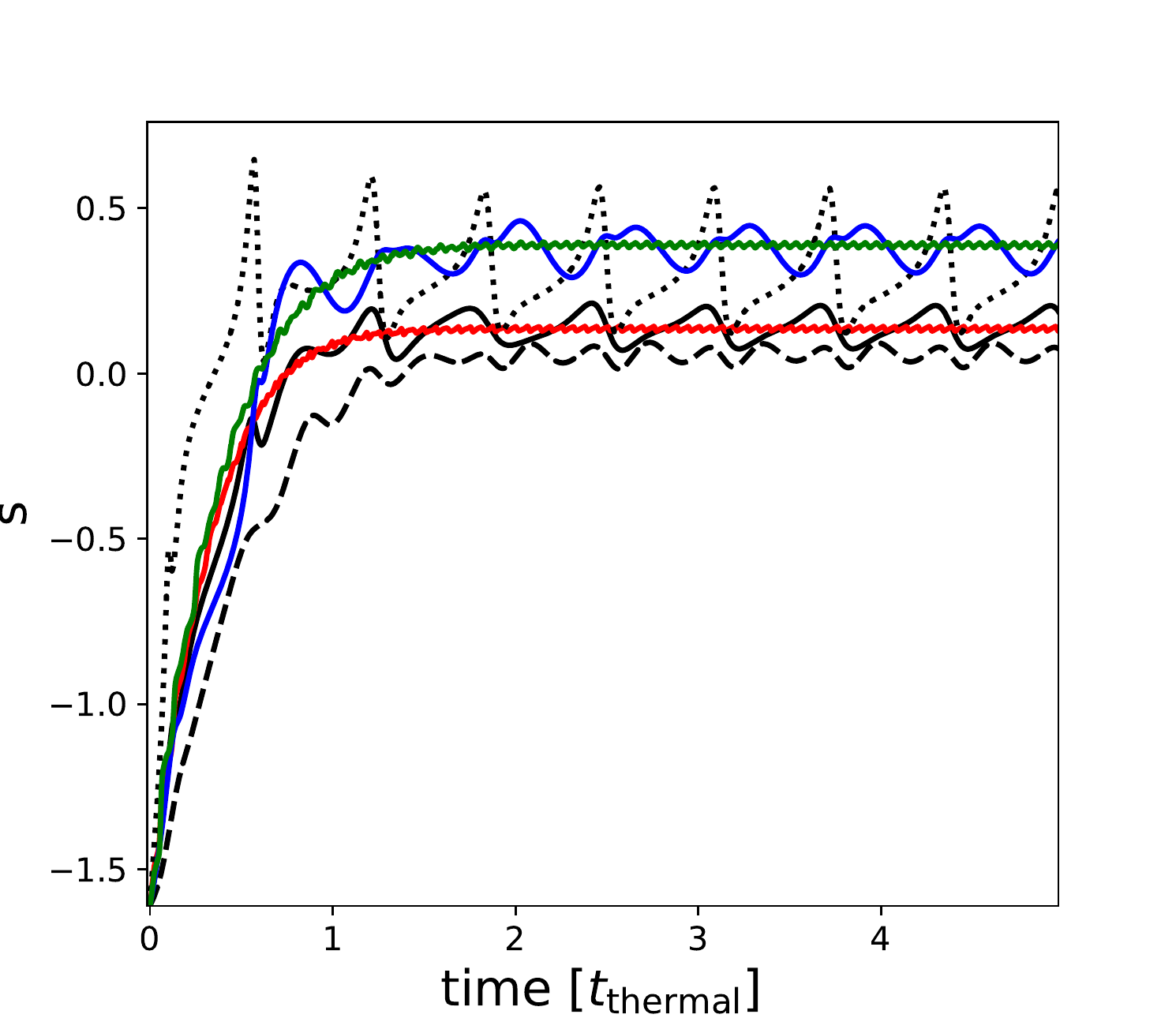}
\caption{Convergence of the specific entropy towards the periodic solution for different disc models. Time is in units of the thermal timescale for the corresponding disc model. Except where noted the disc parameters correspond to those in Figure \ref{example convergence}. Black solid: same as Figure \ref{example convergence}. Red solid: $\alpha_s = 0.01$. Black dashed: $e=0.1$, $q = 0.5$. Black dotted: $e=0.5$, $q=0.1$. Blue solid: $\gamma=1.4$. Green solid: $\gamma=1.4$, $\alpha_s = 0.01$. } 
\label{thermal time damping}
\end{figure}

We find that generically, if the relative errors are controlled, then the solution will relax to a periodic solution on the thermal timescale (see Figure \ref{thermal time damping}), $1/\alpha_s$ in the dimensionless units. The resulting solution is insensitive to the initial guess used; although when $e$ and $q$ are very close to 1 (larger than the $e=q=0.9$ typically considered in this paper) starting far from the periodic solution leads to the solver taking prohibitively short step sizes to control the numerical error.

\section{Scale Height Variation in TDE discs} \label{tde sols}

In this section we study the dynamics of a pure gas ($\Gamma_1 = \Gamma_3 = \gamma$) in a disc with geometry appropriate for an eccentric TDE.

Discs formed by TDEs represent an extreme environment for eccentric disc theory, which forces a very strong variation of the vertical structure around the orbit. In the initial TDE disc the details of the thermodynamics are inseparable from the dynamics of the vertical structure and can be a challenge to numerical methods. \citet{Ogilvie14} showed that for ideal discs the variation of the scale height becomes extreme in the limits $e \rightarrow 1$ or $q \rightarrow 1$ depending on $\gamma$. Both these limits are relevant to TDE discs, with the initial post-disruption orbits having $e \approx q \approx 1$ (see Appendix \ref{e q deriv}). This results in extreme compression of the disc near pericentre in the ideal theory \citep{Ogilvie14}. It might be expected that viscous heating raises the pressure at pericentre and prevents the extreme compression seen in the ideal models; the reality is rather different.

We find that there are two limiting behaviours in these extreme discs, which are compared in Table \ref{dominated vs suppressed}. A pericentre-dominated limit, where the entropy in the disc is strongly peaked at pericentre, and a pericentre-suppressed limit, where the entropy peak at pericentre is much smaller and comparable to the entropy in the rest of the disc. These two limits have some similarities with the two limiting behaviours of the ideal disc. The pericentre-dominated disc typically occurs for low $\gamma$, while the pericentre suppressed disc typically occurs when $\gamma$ is high. 

Between these two extremes there is an intermediate behaviour where the peak at pericentre is comparable to the pericentre-dominated disc but is much broader and extends over a much larger fraction of the orbit. This typically occurs for discs with low $\gamma$ but relatively large $\alpha_s$.

The controlling factor that dictates which behaviour the disc has appears to be the strength of the viscous stresses associated with the vertical compression (i.e. the terms proportional to $\dot{H}$ in Equation \ref{scale height equation}). In the pericentre-dominated limit this stress dominates over the pressure forces and horizontal viscous stresses, particularly around pericentre. It is comparable to the other forces in the pericentre-suppressed limit.

\begin{table*}
\begin{tabular}{| p{0.4 \textwidth} | p {0.4 \textwidth} |}
Pericentre-dominated & Pericentre-suppressed \\
\hline
Characterised by: & Characterised by: \\

Sharp peak in entropy at pericentre which is significantly higher than the rest of the disc & Modest peak in entropy at pericentre that is comparable to the maximum entropy in the rest of the disc \\
\hline
Caused by: & Caused by: \\
\begin{itemize}
\item lower $\gamma$ 
\item higher $e$
\end{itemize} 

&

\begin{itemize}
\item higher $\gamma$ 
\item higher $q$ 
\end{itemize} \\
\hline
Example disc parameters: $e=0.99$, $q=0.9$, $\gamma = 4/3$, $E_0=0$, $\alpha_s = 0.01$ & Example disc parameters: $e=0.9$, $q=0.9$, $\gamma = 5/3$, $E_0=0$, $\alpha_s = 0.1$ \\
\hline
\includegraphics[trim=40 30 40 40,clip,width=\linewidth]{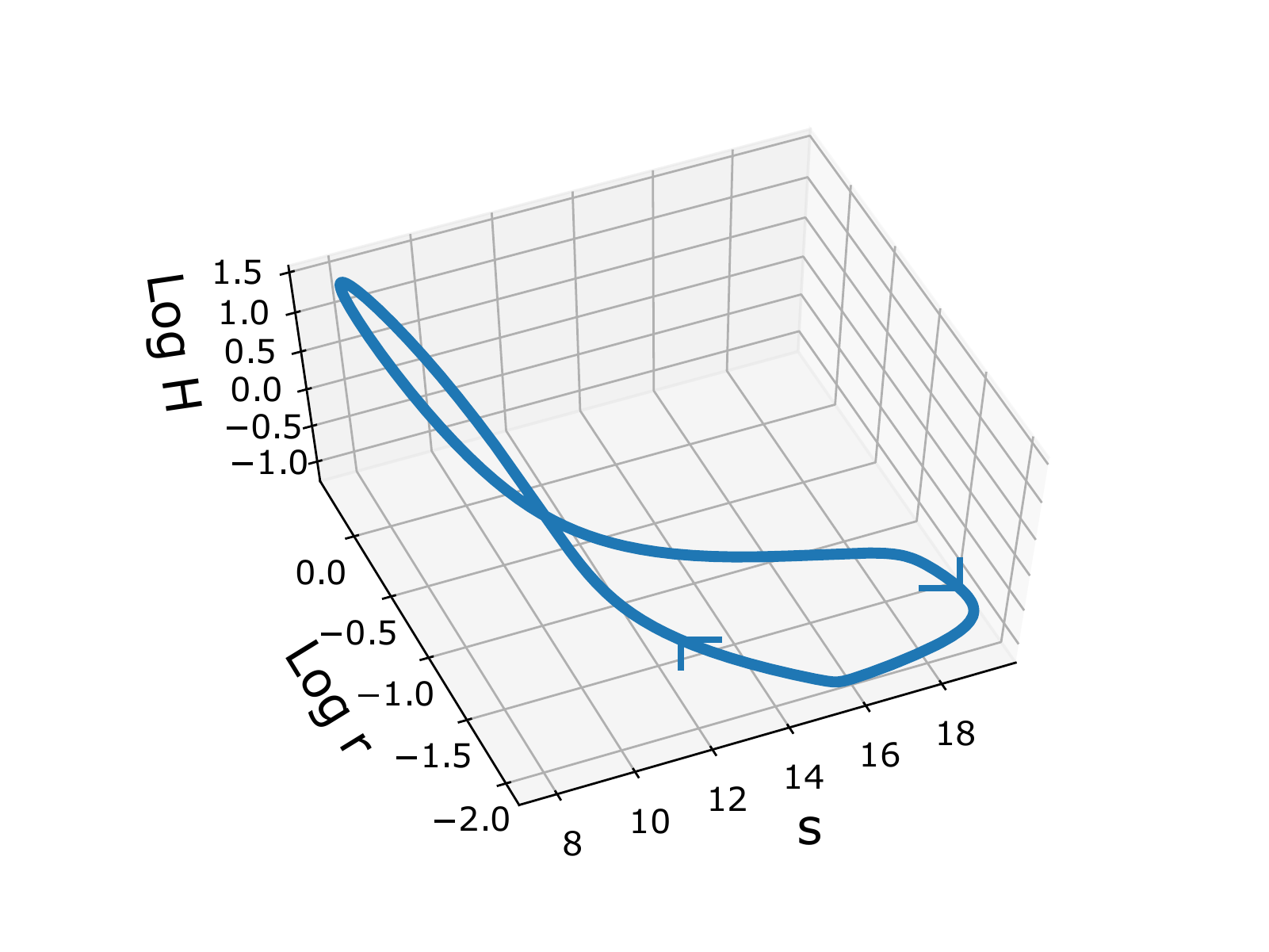}
&
\includegraphics[trim=30 20 40 40,clip,width=\linewidth]{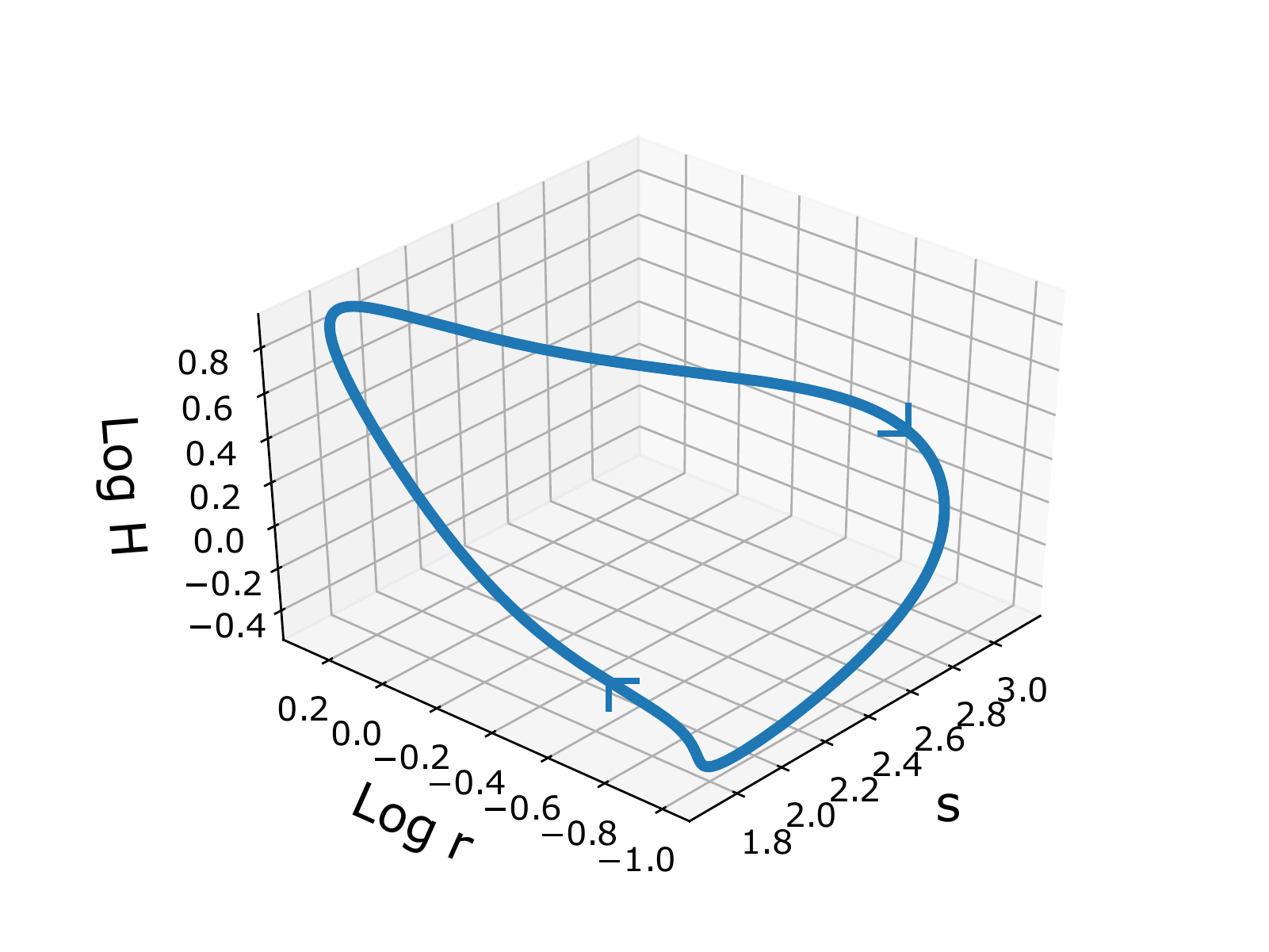}
 \\
\hline 
\includegraphics[width=\linewidth]{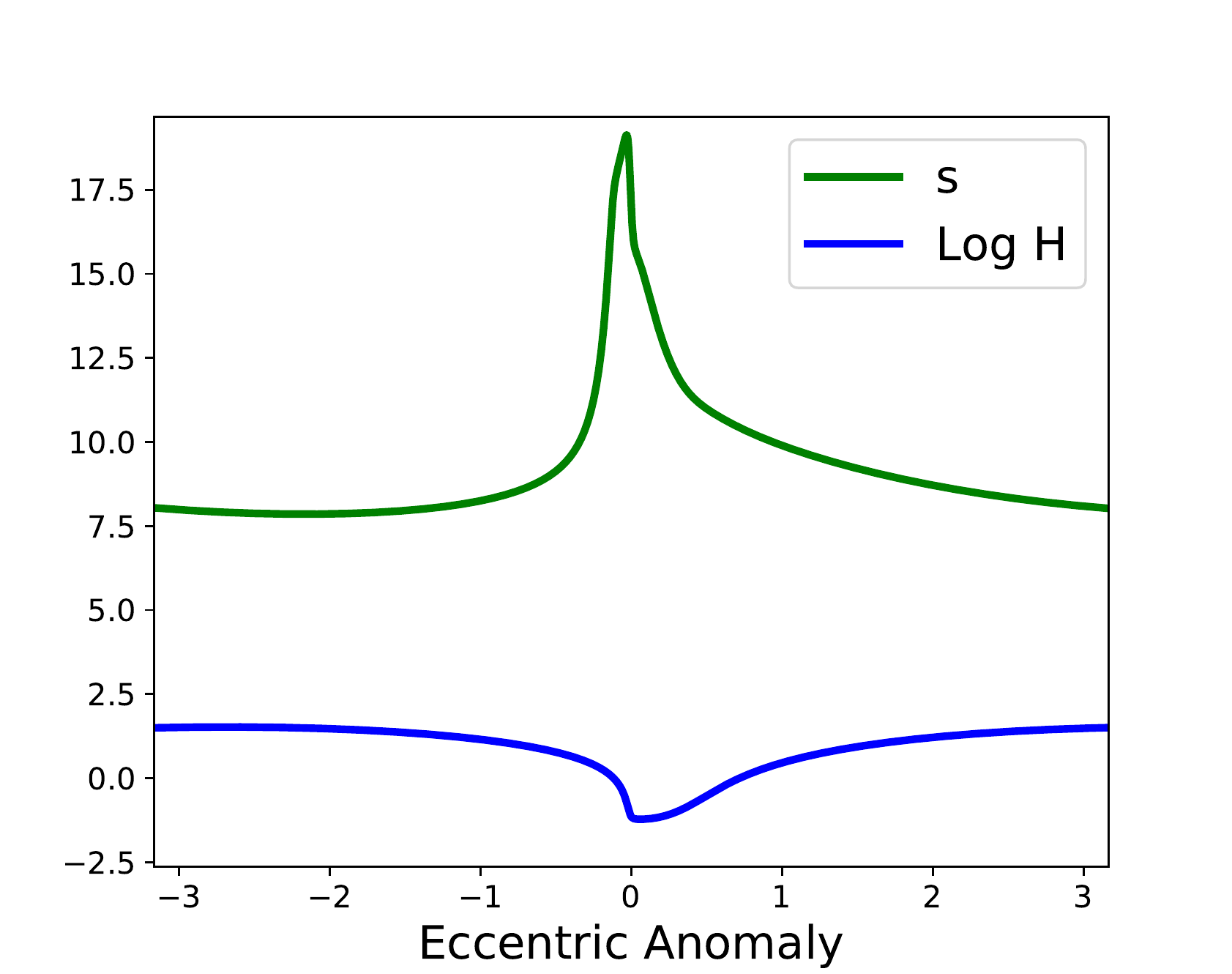}
&
\includegraphics[width=\linewidth]{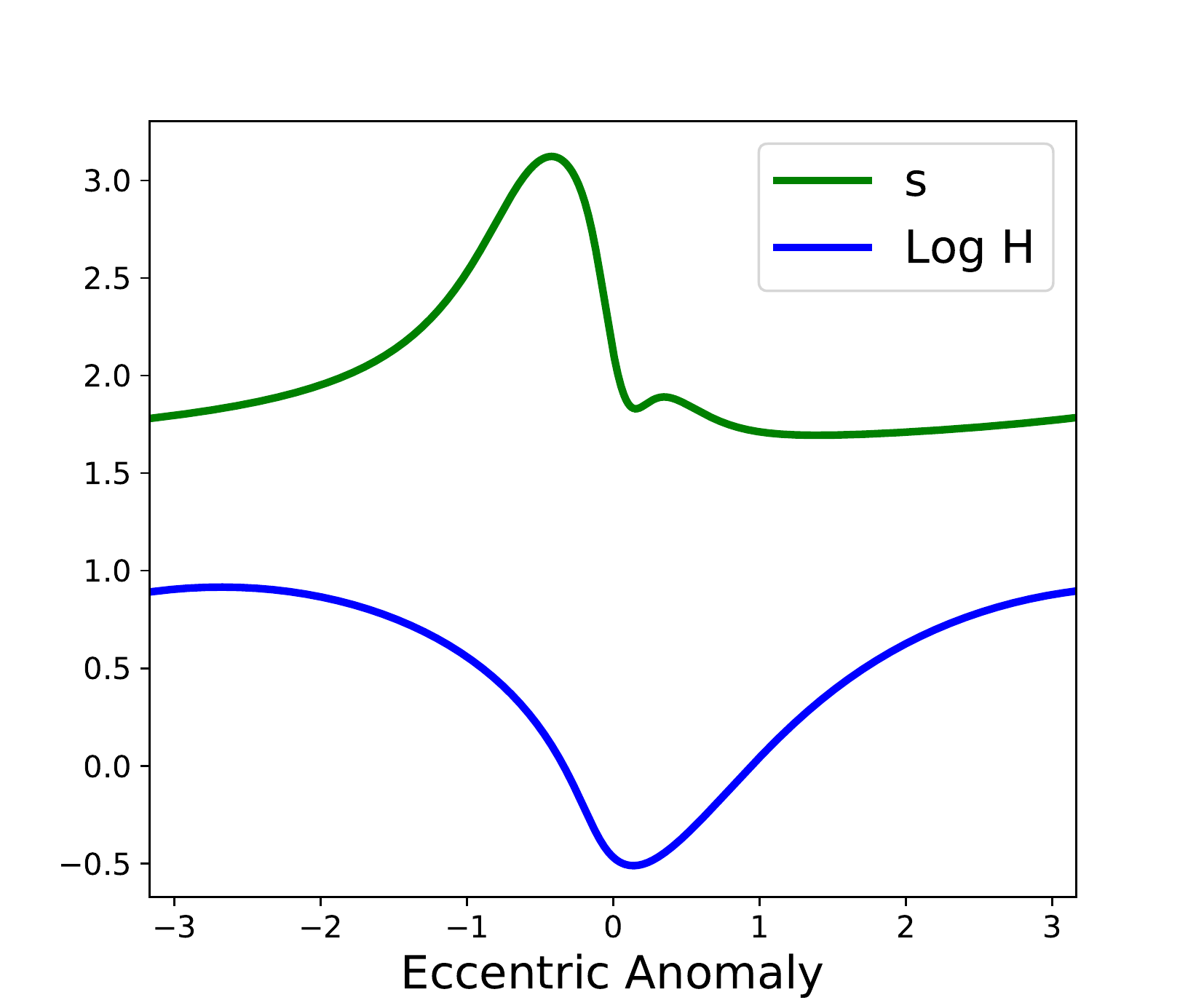}
\\
\hline
Disc showing the pericentre-dominated behaviour. As it returns to pericentre the fluid converges in a vertical ballistic motion, after which it undergoes very strong heating and cooling. In this limit the fluid column can be thought of as collapsing as it approaches pericentre and undergoing abrupt heating and momentum loss causing the fluid to bounce. On exiting pericentre the fluid is accelerated by gas pressure along the receding arc.
&
Disc showing the pericentre-suppressed behaviour. As with the high-$e$ case strong heating occurs at pericentre. However the dominant source of heating comes from horizontal compression along the approaching arc. The fluid forces in the pericentre-suppressed limit are strong enough that the fluid motion on the return trajectory departs strongly from the quasiballistic motion seen in the pericentre-dominated limit. \\
\hline
\end{tabular}
\caption{Comparison of the pericentre-dominated and pericentre-suppressed cases.}
\label{dominated vs suppressed}
\end{table*}

One feature absent in our model is the presence of an orbital intersection at apocentre owing to the strong GR precession.  However TDEs are expected to occur close to the maximum distance capable of disrupting the star \citep{Stone16,Lu19}, for lower mass SMBHs, or higher mass stars, the TDE occurs sufficiently far from the black hole that apsidal precession is weak. In these TDEs the GR precession causes a slow evolution of the disc streamlines rather an intersection of the stream at apocentre, and its main effect on our model is through setting the dependence of $e$ and $\varpi$ on $a$. As we are dealing with a local model centred on one orbit this weak apsidal precession has no direct effect on our solutions. Our model is incapable of dealing with the strong apsidal precession seen in TDEs with pericentric distances comparable to the gravitational radius (e.g. as simulated in \citet{Hayasaki13,Sadowski16}).

In our disc we have no orbital intersection (which would correspond to $q=\pm 1$) and there is no way for the model to be close to orbital intersection simultaneously at pericentre and apocentre without the eccentricity oscillating on a short lengthscale or the disc being highly twisted. In fact with the positive eccentricity gradient expected in a tidal disruption event the disc should only be near to an orbital intersection in the nozzle region at pericentre. For TDEs with smaller impact parameters GR precession will be stronger and we expect this will cause the disc to become increasingly twisted (corresponding to $E_0 \rightarrow \pi$) eventually resulting in an orbital intersection close to apocentre. We leave consideration of disc twist to future work.

\section{Radiation/Gas Mixture} \label{radgas}

We now consider the dynamics of a TDE including both radiation and gas pressure. The disc's vertical structure is assumed to undergo vertical mixing (either from convection or the disc turbulence), which provides an additional source of heat transport that is comparable to the radiative flux near the midplane. It is assumed that the vertical mixing sets up a polytropic vertical structure with polytropic index $3$, which implies that $\beta_r$ is independent of height. In the radiation pressure dominated limit this leads to a vertically isentropic structure. This assumption is required for the equations to remain separable, although it has no other effect on the variation of the scale height around the orbit. $\Gamma_3$ is now given by Equation \ref{Gamma 3 def} and we solve Equations \ref{scale height equation} and \ref{thermal energy equation T}.

\subsection{Stress scaling with total pressure ($p_v = p$)}

\begin{figure}
\includegraphics[trim=0 0 0 0,clip,width=\linewidth]{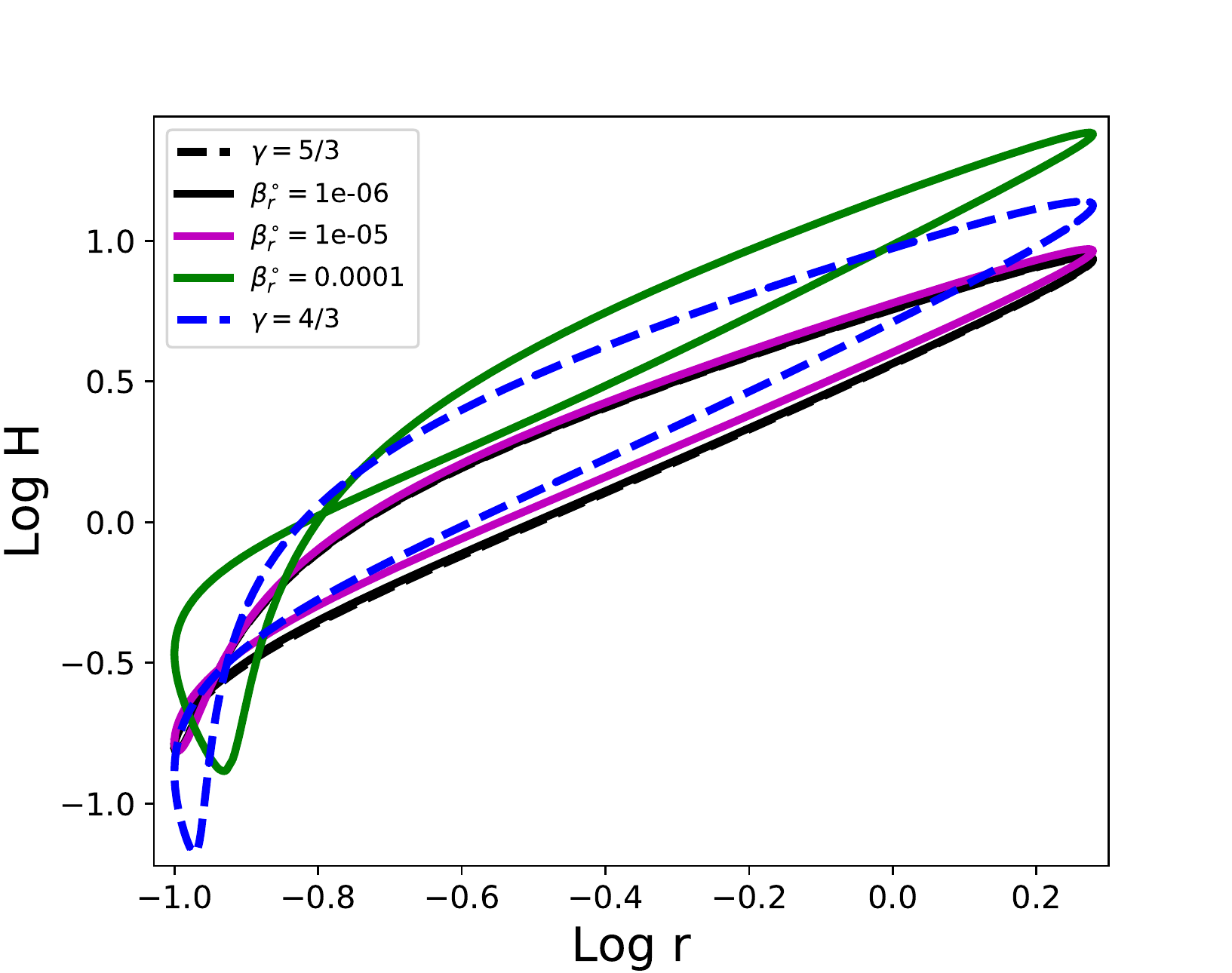}
\caption{Variation of the scale height of the disc for a radiation-gas mixture with $p_v = p$. The dashed lines are for pure gas discs with $\gamma$ of $5/3$ and $4/3$. Disc parameters are $\alpha_s=0.01$, $\alpha_b=0$, $e=q=0.9$ and $E_0 = 0$. For low $\beta_{r}^{\circ}$ the disc behaves like the pure gas disc with $\gamma = 5/3$. Discs with larger $\beta_{r}^{\circ}$ are unstable to the thermal instability.} 
\label{radmix model comp}
\end{figure}

\begin{figure}
\includegraphics[trim=0 0 0 0,clip,width=\linewidth]{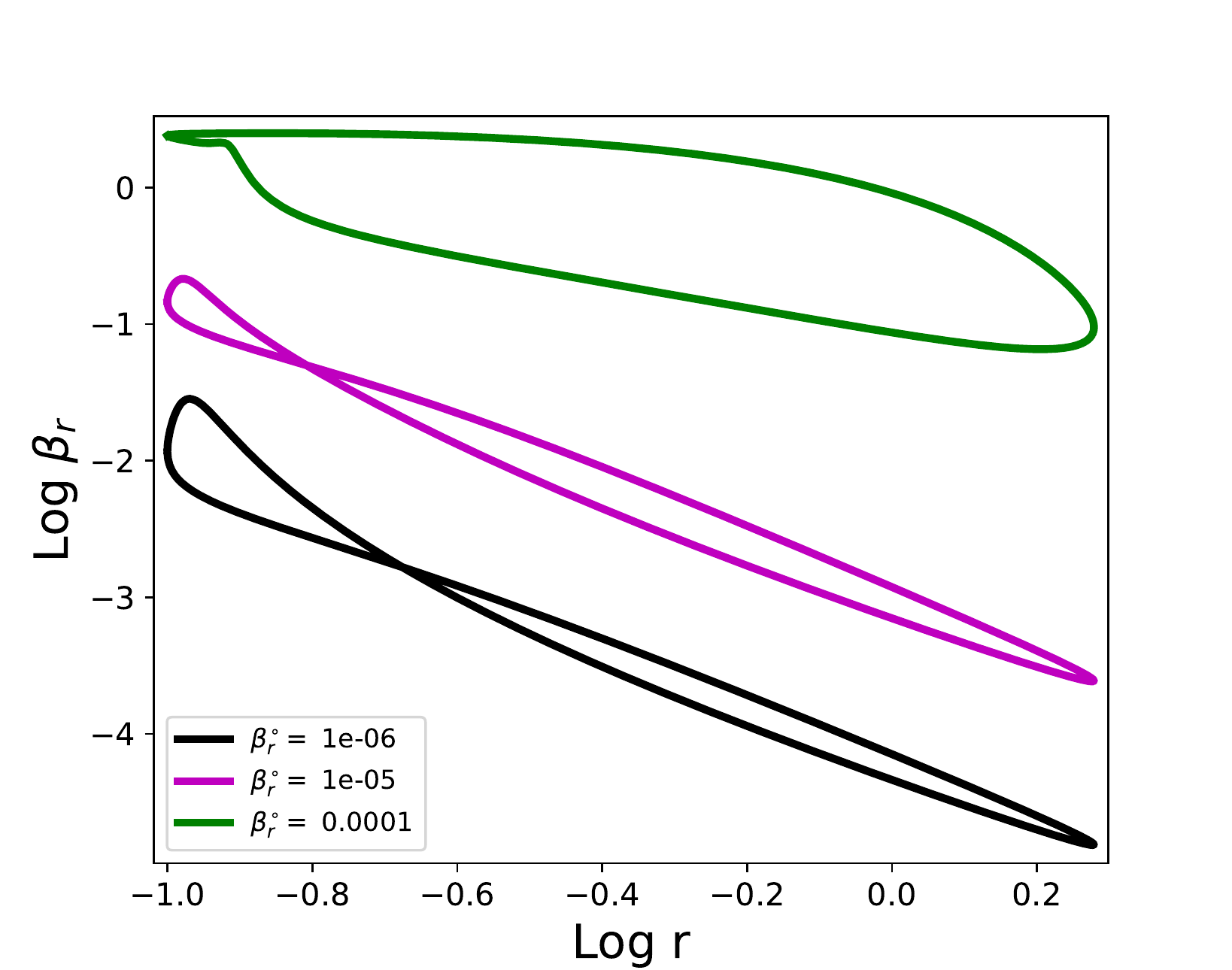}
\caption{Variation of $\beta_r$ around the orbit for each model in Figure \ref{radmix model comp}. Radiation pressure becomes most important in the  compression at pericentre.}  
\label{radmix model beta}
\end{figure}

\begin{figure}
\includegraphics[trim=0 0 0 0,clip,width=\linewidth]{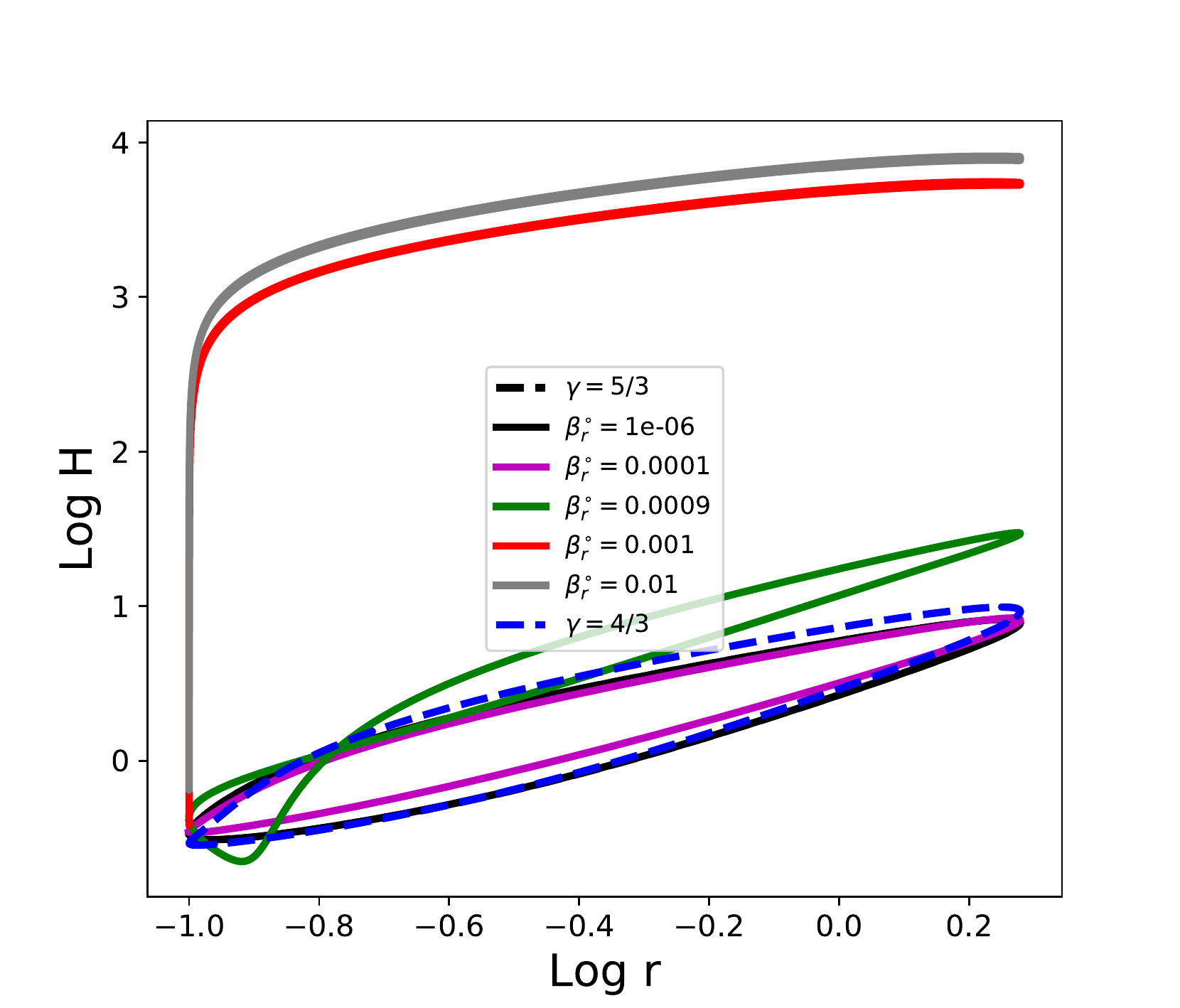}
\caption{Same as Figure \ref{radmix model comp} but with $p_v = p_g$ and $\alpha_s = 0.1$. Much larger $\beta_r$ can be achieved owing to the stabilisation of the thermal instability. There appears to be an abrupt change in behaviour between $\beta_{r}^{\circ} = 9 \times 10^{-4}$ and $\beta_{r}^{\circ} = 1 \times 10^{-3}$ where the disc becomes approximately adiabatic.} 
\label{highbeta radmix model comp}
\end{figure}

\begin{figure}
\includegraphics[trim=0 0 0 0,clip,width=\linewidth]{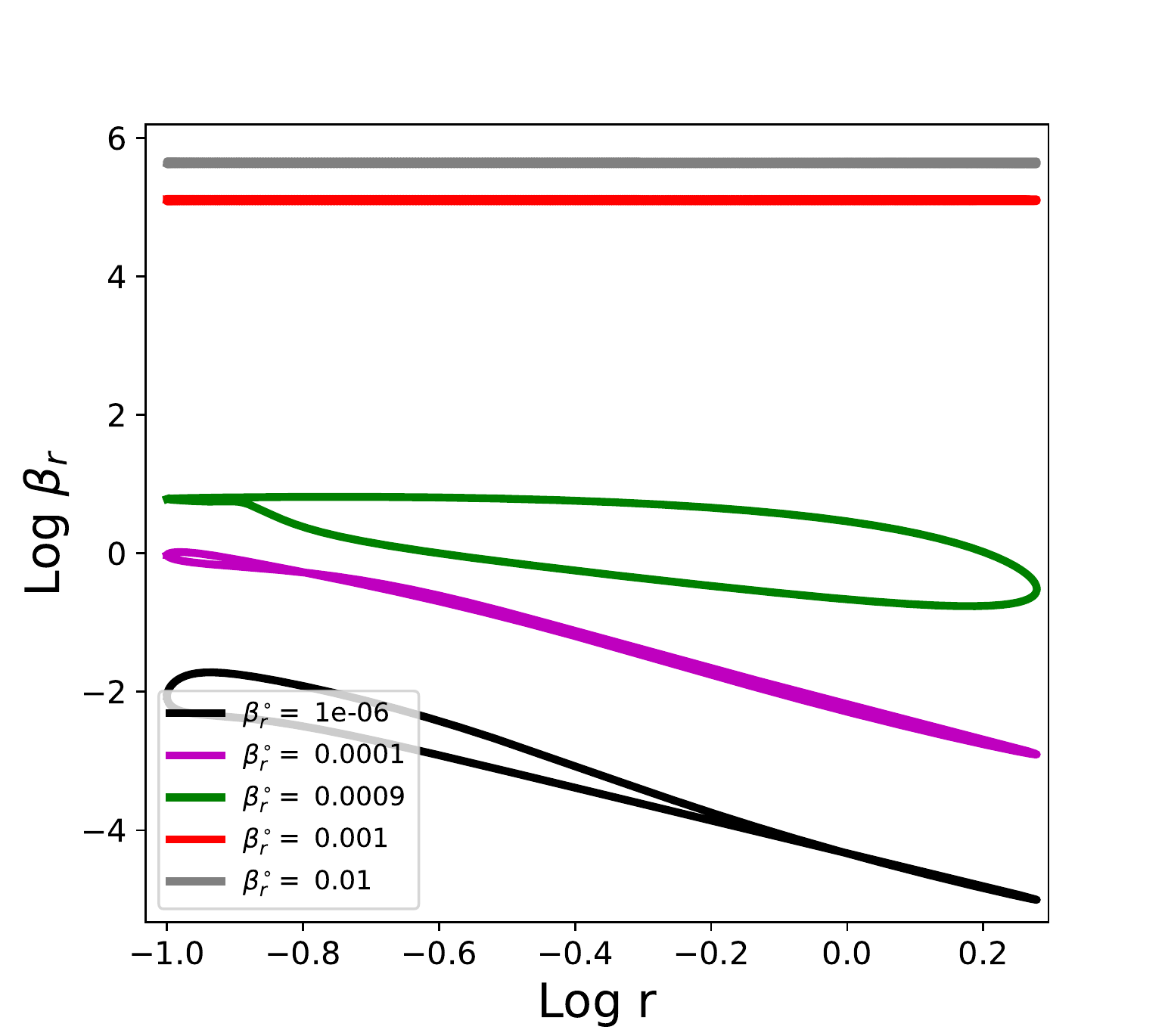}
\caption{Variation of $\beta_r$ around the orbit for the models in Figure \ref{highbeta radmix model comp}. We can now attain $\beta_{r} \gg 1$. In the radiation dominated limit a nearly constant $\beta_r$ implies a nearly  constant entropy (see Equation \ref{kentrop radmix}).}  
\label{highbeta radmix model beta}
\end{figure}

Figures \ref{radmix model comp} and \ref{radmix model beta} show how the scale height and $\beta_r$ vary around the orbit for different $\beta_{r}^{\circ} $ for a disc with $e=q=0.9$, $\alpha_s = 0.01$ and $E_0 = \alpha_b = 0$ and stress scaling with total pressure ($p_v = p$). In general the $\beta_r$ of the disc can exceed $\beta_r^{\circ}$ by several orders of magnitude as the physical conditions in the eccentric disc are very different from those of the circular reference state, primarily due to the enhanced heating rates occurring at pericentre.

The importance of radiation pressure is greatest in the strong compression experienced during pericentre passage. However even at pericentre $\beta_{r}$ cannot significantly exceed $1$ without triggering the thermal instability. This is distinct from the behaviour of the pure gas with $\gamma=4/3$, normally used to mimic a radiation-dominated disc, as we find this to be thermally stable. This highlights the difference between a $\gamma=4/3$ gas and a radiation-dominated disc when heating and cooling are present. The opposite limit $\beta_{r}^{\circ} \rightarrow 0$ matches the behaviour of the pure gas solutions with $\gamma = 5/3$.

The initial TDE disc tends to be highly radiation pressure dominated \citep{Loeb97,Shen14}, for instance the estimate in \citet{Loeb97} gives  $\beta_{r} \approx 10^4$. Similarly Zanazzi \& Ogilvie (2020, submitted to MNRAS) considered TDEs with $\beta_{r} \approx 10^4$ while \citet{Bonnerot19} found $\beta_{r}$ of $10^5$. While we can slightly increase $\beta_{r}$ by reducing $\alpha_s$, in general our models are incapable of reaching as high $\beta_r$ as the values quoted, owing to the presence of the thermal instability.  In our model, thermally unstable circular discs can collapse to a thin gas pressure dominated state or experience thermal runaway with the temperature and disc thickness increasing with each orbit.  Our thermally unstable highly eccentric discs almost universally experience thermal runaway.

\subsection{Stress scaling with gas pressure ($p_v = p_g$)}

In order to reach realistic $\beta_r$ we test solutions where the stress scales with gas pressure ($p_v = p_g$). Figures \ref{highbeta radmix model comp} and \ref{highbeta radmix model beta} show how the scale height and $\beta_r$ vary around the orbit for different $\beta_{r}^{\circ}$ for this case. The disc parameters are $e=q=0.9$ and $E_0 = \alpha_b = 0$ like those considered in Figures \ref{radmix model comp} and \ref{radmix model beta}; however we set $\alpha_s = 0.1$. These can now attain realistic values of $\beta_r$ without encountering an unbounded thermal runaway. 

When $\beta_{r}^{\circ}$ is sufficiently high the disc becomes strongly radiation pressure dominated and behaves nearly adiabatically. The transition to this behaviour as we vary $\beta_{r}^{\circ}$ is quite abrupt and may represent a change in solution branch.  Our interpretation of this result is that, for highly eccentric discs with $p_v = p_g$, there are two solution branches: a gas pressure dominated solution branch with significant entropy variation and a nearly adiabatic radiation pressure dominated solution branch. When $\beta_{r}^{\circ}$ is sufficiently high the gas pressure dominated branch is thermally unstable and the solution switches to the other branch. This nearly adiabatic radiation pressure dominated branch seems to be peculiar to eccentric discs as the circular disc has only one solution branch, which is stable regardless of $\beta_{r}^{\circ}$. When stress scales with total pressure the nearly adiabatic radiation pressure dominated branch doesn't appear to exist (or if it does it is always unstable) as the solution undergoes an unbounded thermal runaway.

We find that for our radiation pressure dominated discs there is a long lived transient state in which the solution is well approximated by an adiabatic solution on a orbital timescale, but undergoes a slow drift in (orbit-averaged) entropy over many orbits. This is related to the freedom to specify the disc entropy in a strictly ideal disc. Eventually the solution settles down to a fixed orbit-averaged entropy; there is still entropy variation around the orbit but this is small compared with the orbit-averaged entropy and has little effect on the dynamics. 

The radiation dominated solution and the $\gamma=4/3$ perfect gas with $\alpha_s \ll 1$ go through the same nearly adiabatic transient entropy drift phase before converging on a specific value for the orbit-averaged entropy. However, owing to the difference in the entropy/temperature for a $\gamma=4/3$ gas and a radiation dominated disc, these converge to different nearly adiabatic solutions! This leads to a disagreement between eccentric radiation pressure dominated discs and eccentric $\gamma=4/3$ discs, which we shall explore further in Section \ref{rad gas perf gas comp}.

Both the radiation dominated solution and $\gamma=4/3$ adiabatic (or $\alpha \ll 1$) solution exhibit extreme compression near pericentre (see also \citet{Ogilvie14}). The reader might be concerned that the $H$ can reach $10^3-10^4$ at apocentre and appears to strongly violate the thin-disc assumptions. However, the disc scale height is normalised with respect to the scale height of a reference circular disc, which can be very thin. For a typical $1 M_{\odot}$ TDE around a $10^6 M_{\odot}$ black hole, however, $H^{\circ}/a \sim 10^{-3}$ (see Appendix \ref{e q deriv}), so $H/r$ is $\sim 1$ at apocentre and the thin disc assumptions are violated. These discs start out geometrically thin around the entire orbit, prior to the onset of thermal instability. In general, the onset of thermal instability can be treated by our model, while the outcome, even for $p_v=p_g$, cannot. It is likely that additional physics, such as a coherent magnetic field, will reduce or prevent the extreme behaviour seen in the radiation dominated models. We shall explore this possibility further in a future paper.

If we increase $\alpha_s$ it is possible to obtain radiation pressure dominated solutions which exhibit significant entropy variation around the orbit. In these solutions viscosity and radiative cooling are again important and they have a similar behaviour to the marginally gas pressure dominated solutions in the models where stress scales with total pressure (but attain higher $\beta_{r}$ at pericentre). However, these solutions typically require $\alpha_s > 1$ and it's not clear that this is physically reasonable, although this $\alpha_s$ is defined with respect to the gas pressure. 

\begin{figure}
\includegraphics[trim=10 0 10 10,clip,width=\linewidth]{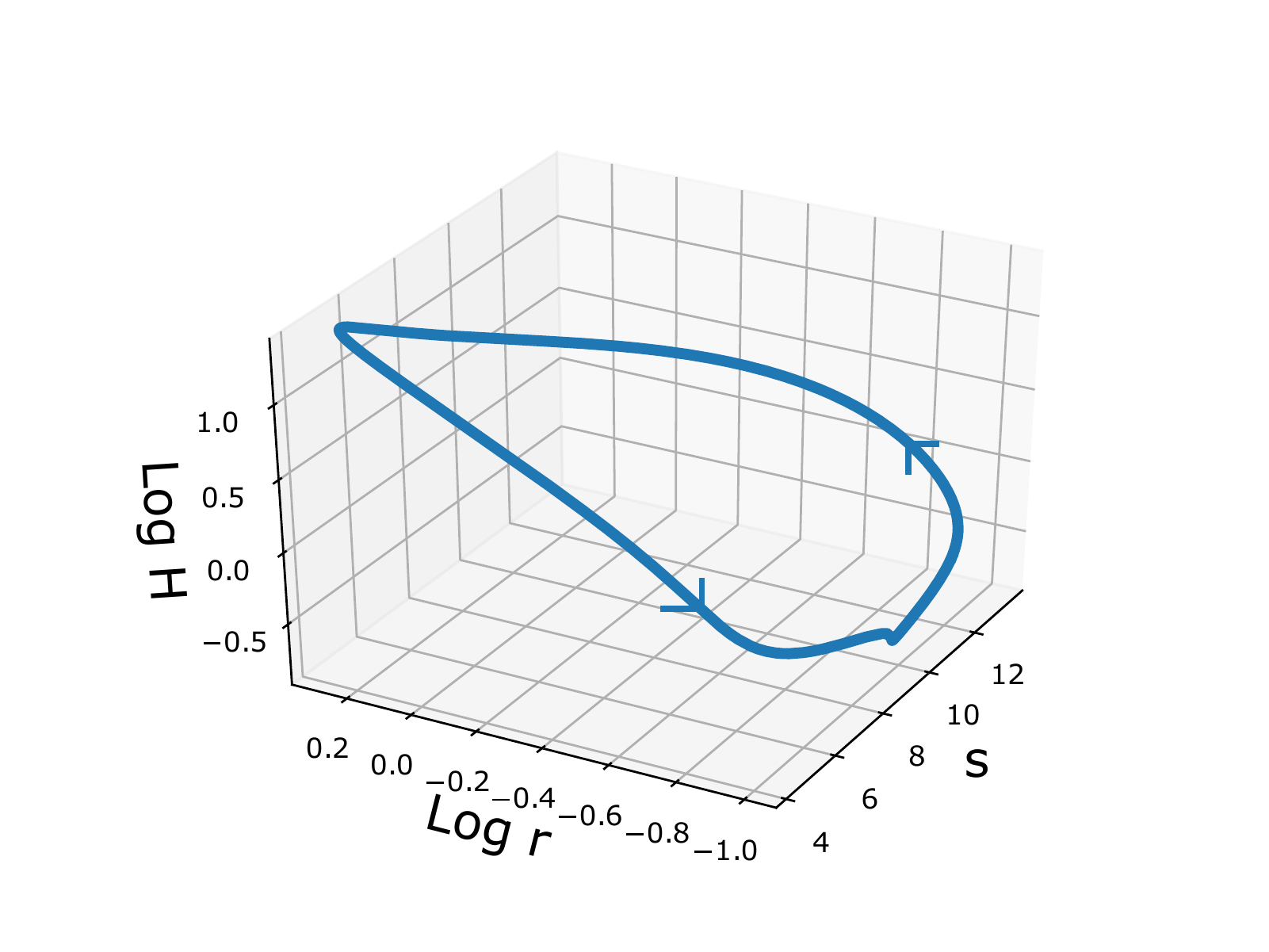}
\caption{Disc with radiation-gas mixture with $p_v = p$,  $\beta_{r}^{\circ}  = 10^{-4}$, $\alpha_s = 0.01$, $\alpha_b = 0$, $e=q=0.9$ and  $E_0 = 0$.} 
\label{radmix traj}
\end{figure}

\begin{figure}
\includegraphics[trim=10 0 0 10,clip,width=\linewidth]{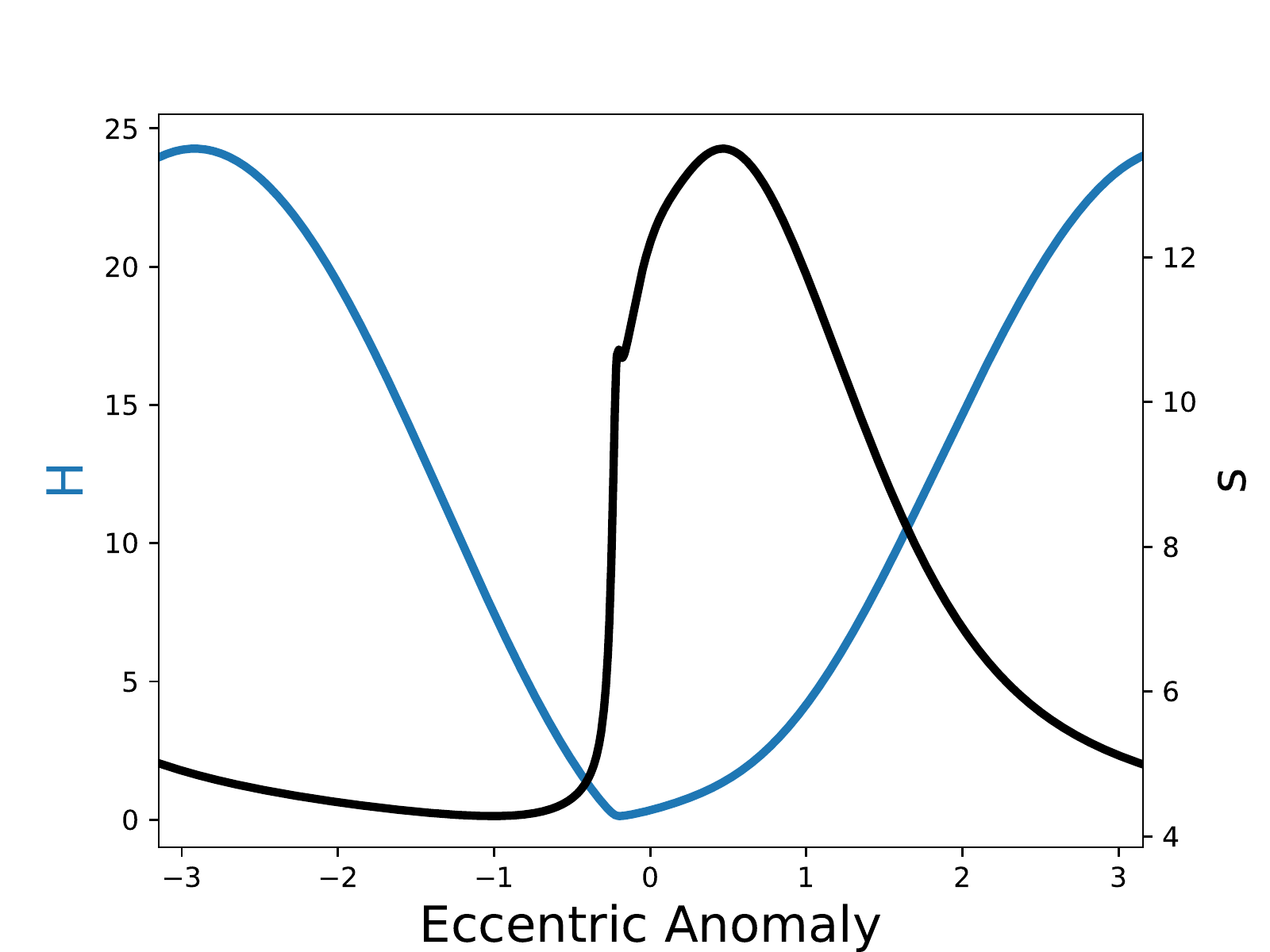}
\caption{Same as Figure \ref{radmix traj} but showing how $H$ and $s$ vary with $E$. The entropy varies strongly around the orbit as in the $\gamma=4/3$ disc; however, after an abrupt increase in entropy during pericentre passage, the entropy declines gradually over half an orbit.} 
\label{radmix eccplot}
\end{figure}

\begin{figure}
\includegraphics[trim=10 30 20 20,clip,width=\linewidth]{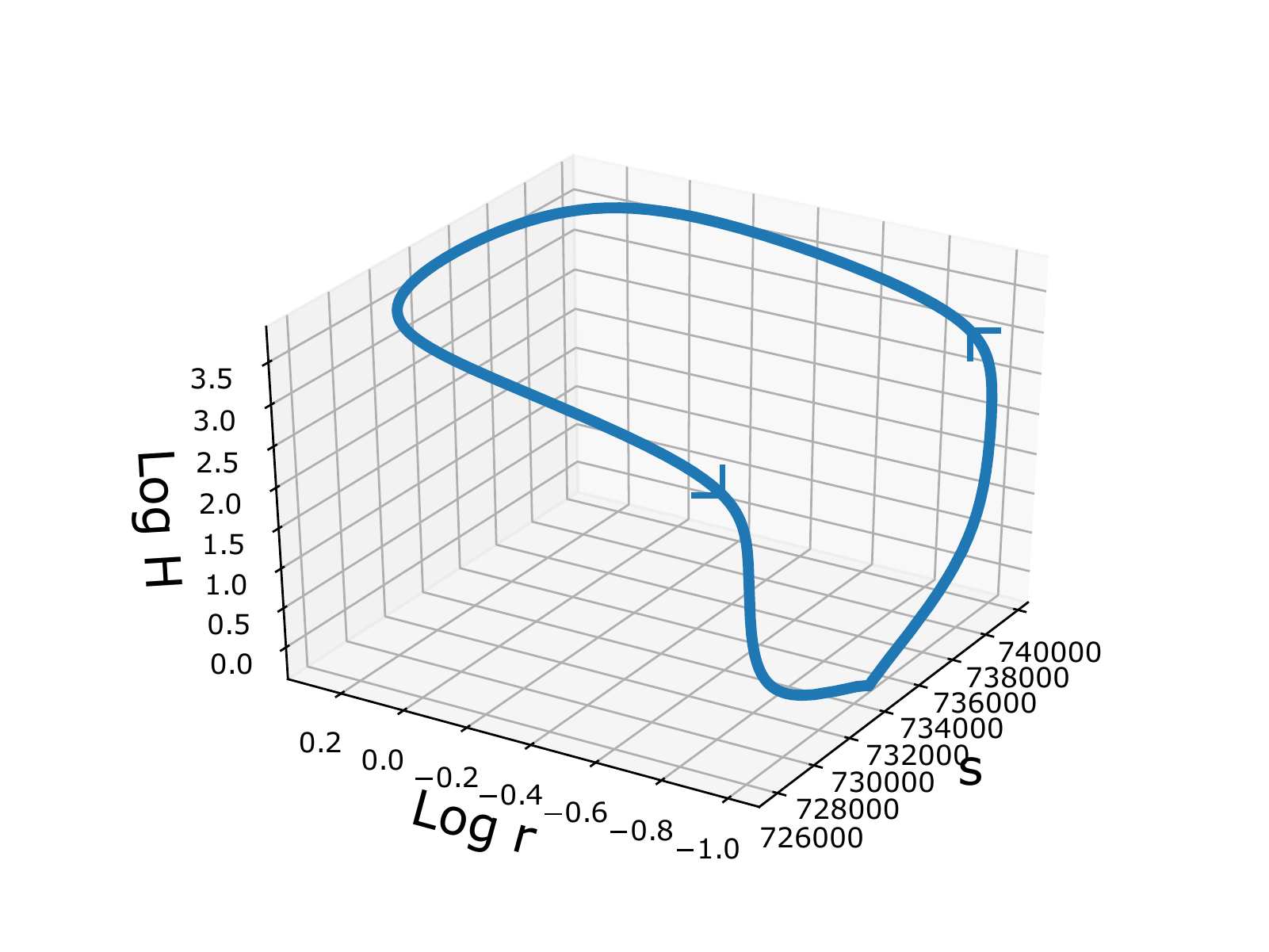}
\caption{Disc with radiation-gas mixture with $p_v = p_g$, $\beta_{r}^{\circ}  = 0.001$, $\alpha_s=0.1$, $\alpha_b = 0$, $e=q=0.9$,  and $E_0 = 0$.}  
\label{highbeta radmix traj}
\end{figure}

\begin{figure}
\includegraphics[trim=20 0 0 0,clip,width=\linewidth]{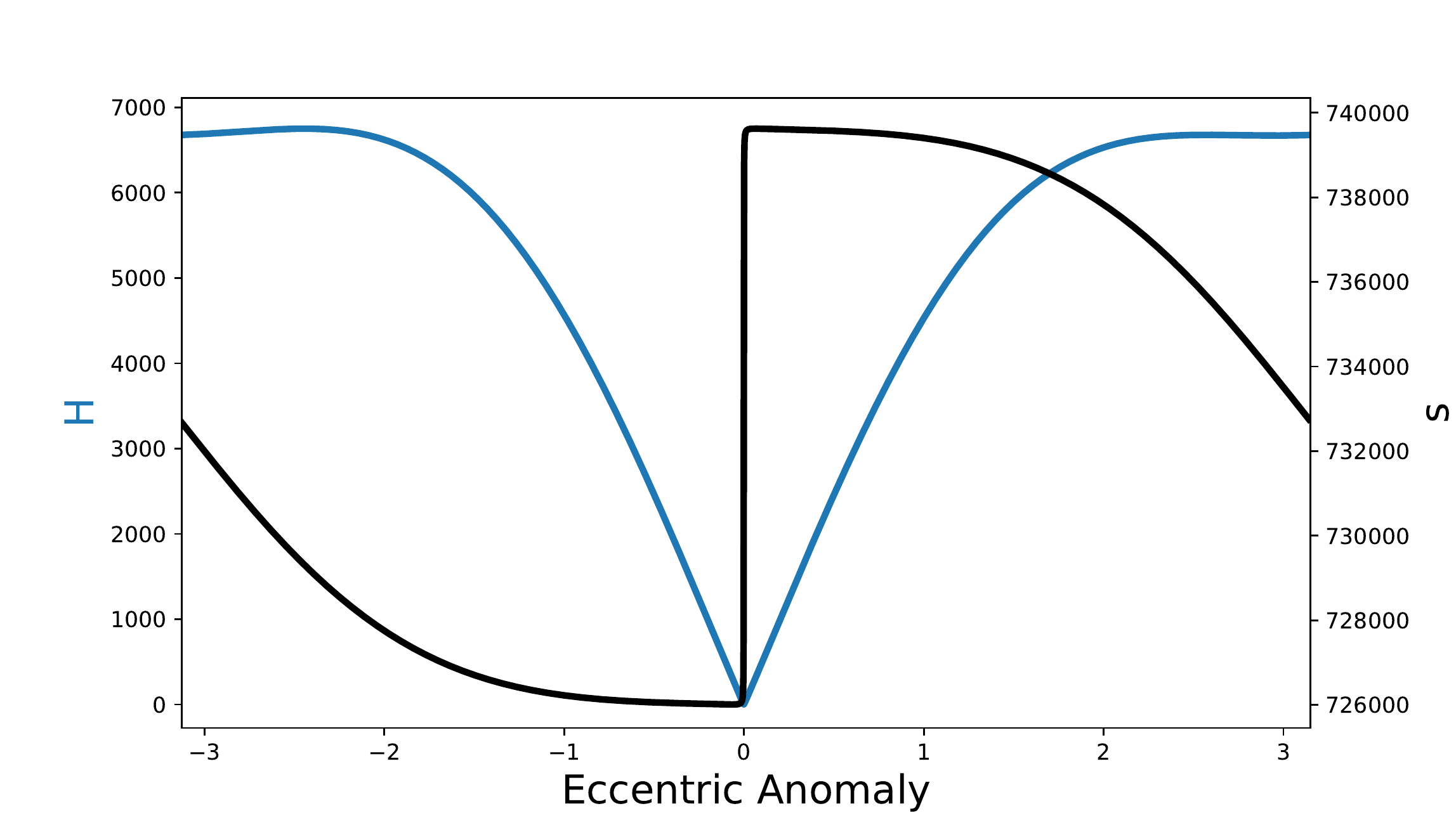}
\caption{Same as Figure \ref{highbeta radmix traj} but showing how $H$ and $s$ vary with $E$. The entropy shows very similar features to that seen in Figure \ref{radmix eccplot}, in particular the abrupt jump at pericentre, however the relative change in entropy around the orbit is small.  The scale height is almost symmetric about pericentre indicating a lack of dissipation.} 
\label{highbeta radmix eccplot}
\end{figure}

Figures \ref{radmix traj} and \ref{radmix eccplot} shows the solution for one of the most radiation dominated discs we can achieve when stress scales with total pressure (with disc parameters $\beta_{r}^{\circ}  = 10^{-4}$, $\alpha_s = 0.01$, $\alpha_b = 0$, $e=q=0.9$ and  $E_0 = 0$). Similarly Figures \ref{highbeta radmix traj} and \ref{highbeta radmix eccplot} show a radiation dominated disc when stress scales with gas pressure (with disc parameters  $\beta_{r}^{\circ}  = 0.001$, $\alpha_s=0.1$, $\alpha_b = 0$, $e=q=0.9$,  and $E_0 = 0$). Both exhibit abrupt changes in the entropy near pericentre, coincident with a reversal in the direction of vertical motion; we examine this behaviour in more detail in Section \ref{nozzle} below. The variation in entropy in the latter solution is small compared to the total entropy and this solution is close to adiabatic.

\begin{figure}
\includegraphics[trim=0 0 0 0,clip,width=\linewidth]{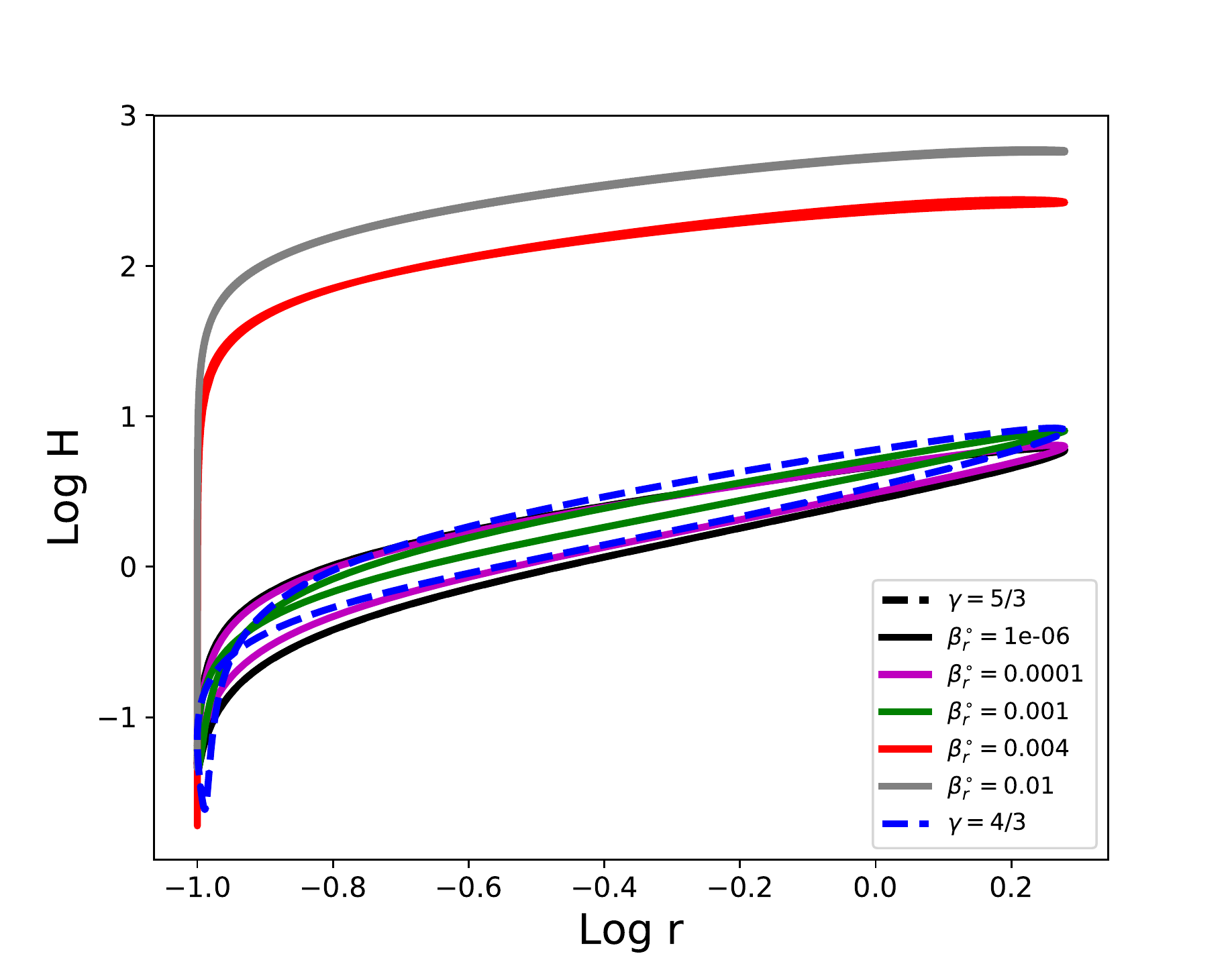}
\caption{Same as Figure \ref{highbeta radmix model comp} but with $\omega_{\rm orb} = \Omega$, showing similar physical behaviour with the alternative choice of frequency. The onset of the thermal instability for this model occurs at higher $\beta_r^{\circ}$.} 
\label{radmix model comp alpha mod}
\end{figure}

\begin{figure}
\includegraphics[trim=0 0 0 0,clip,width=\linewidth]{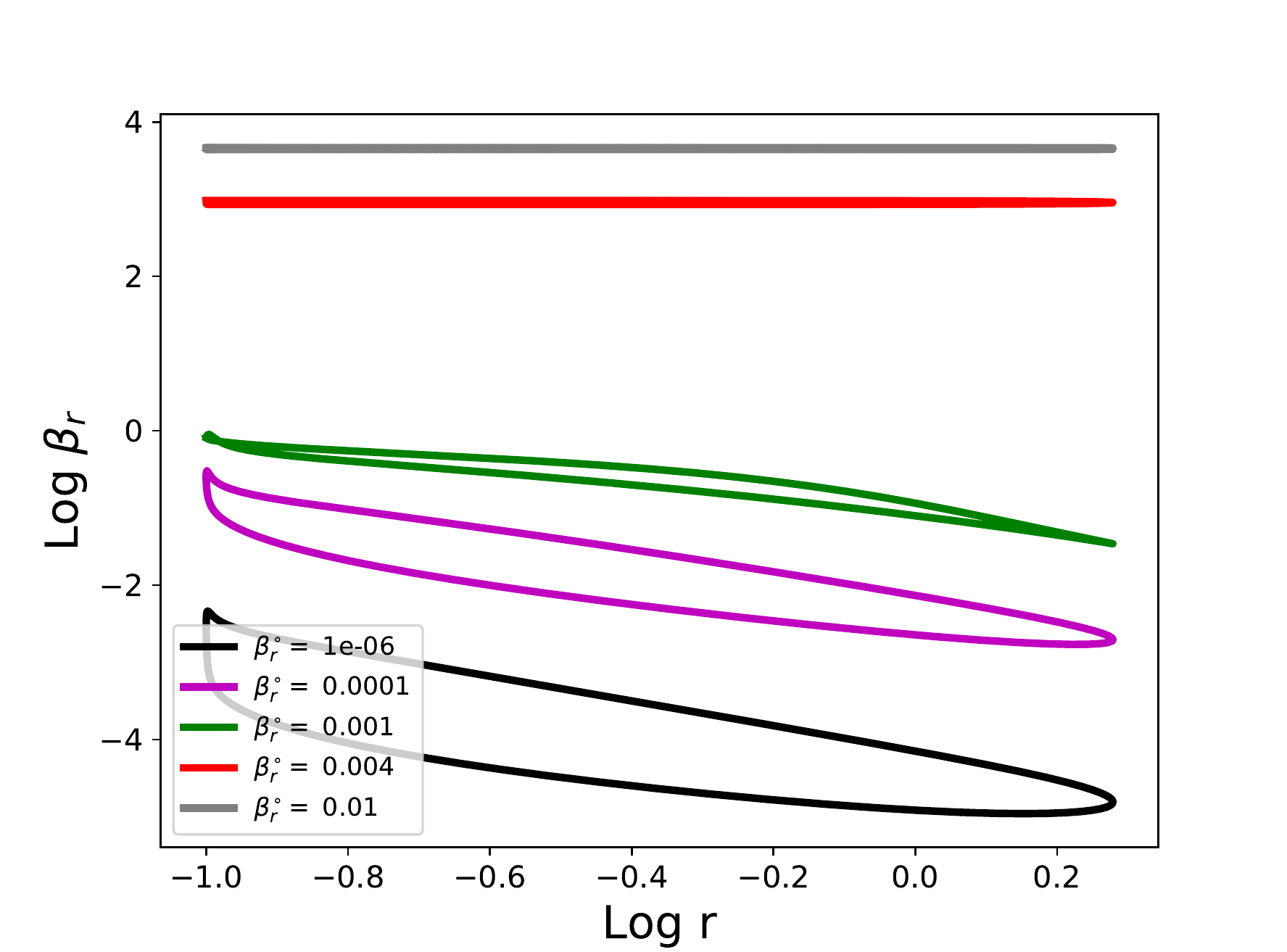}
\caption{Variation of $\beta_r$ around the orbit for the models in Figure \ref{radmix model comp alpha mod}.}  
\label{radmix model beta alpha mod}
\end{figure}

Finally Figures \ref{radmix model comp alpha mod} and \ref{radmix model beta alpha mod} show examples of solutions with a different choice for the orbital frequency $\omega_{\rm orb}$ in the $\alpha-$prescription (here $\omega_{\rm orb} = \Omega$). The solutions have broadly the same physical behaviour as those with $\omega_{\rm orb} = n$, with the thermal instability requiring larger $\beta_r^{\circ}$ (although it can still occur for gas pressure dominated reference discs). These solutions have a more pronounced compression at pericentre than their $\omega_{\rm orb} = n$ counterparts, which appears to be characteristic of discs with weaker viscosities. Adopting a slightly larger value for $\alpha_s$ removes this difference.

\subsection{Implications for TDE thermodynamics}

So what can we conclude about the disc thermodynamics? When stress is proportional to total pressure (e.g. Figures \ref{radmix traj} and \ref{radmix eccplot}) we have strong variation of the entropy around the orbit so our model is not well described by an adiabatic or isothermal disc. The thermal physics of our disc is quite different from the simplified thermodynamic models considered in \citet{Hayasaki16} and \citet{Bonnerot16}. The $\gamma=4/3$ pure gas model is not a good model of a radiation pressure dominated disc when stress scales with total pressure. In particular we find that the radiation dominated discs are thermally unstable, while the $\gamma=4/3$ disc is stable. It is likely that the simulations of TDEs do not encounter the thermal instability owing to their simplified treatment of thermodynamics.

For our radiation dominated solutions (with stress scaling with gas pressure) the relative entropy variation around the orbit is small and, on an orbital timescale, the solution can be well approximated by an adiabatic solution. However this apparent adiabaticity is only maintained because the oscillator is extracting energy from the orbital motion to make up for radiative losses. On longer timescales the orbits must evolve due to radiative losses, which will cause the entropy of the solution to adjust. That the entropy of our radiation dominated solutions can adjust is attested by the fact that the entropy grows by many orders of magnitude (from the starting $s$ close to 1) as the solution relaxes to a periodic state.

\begin{figure}
\includegraphics[trim=0 0 0 0,clip,width=\linewidth]{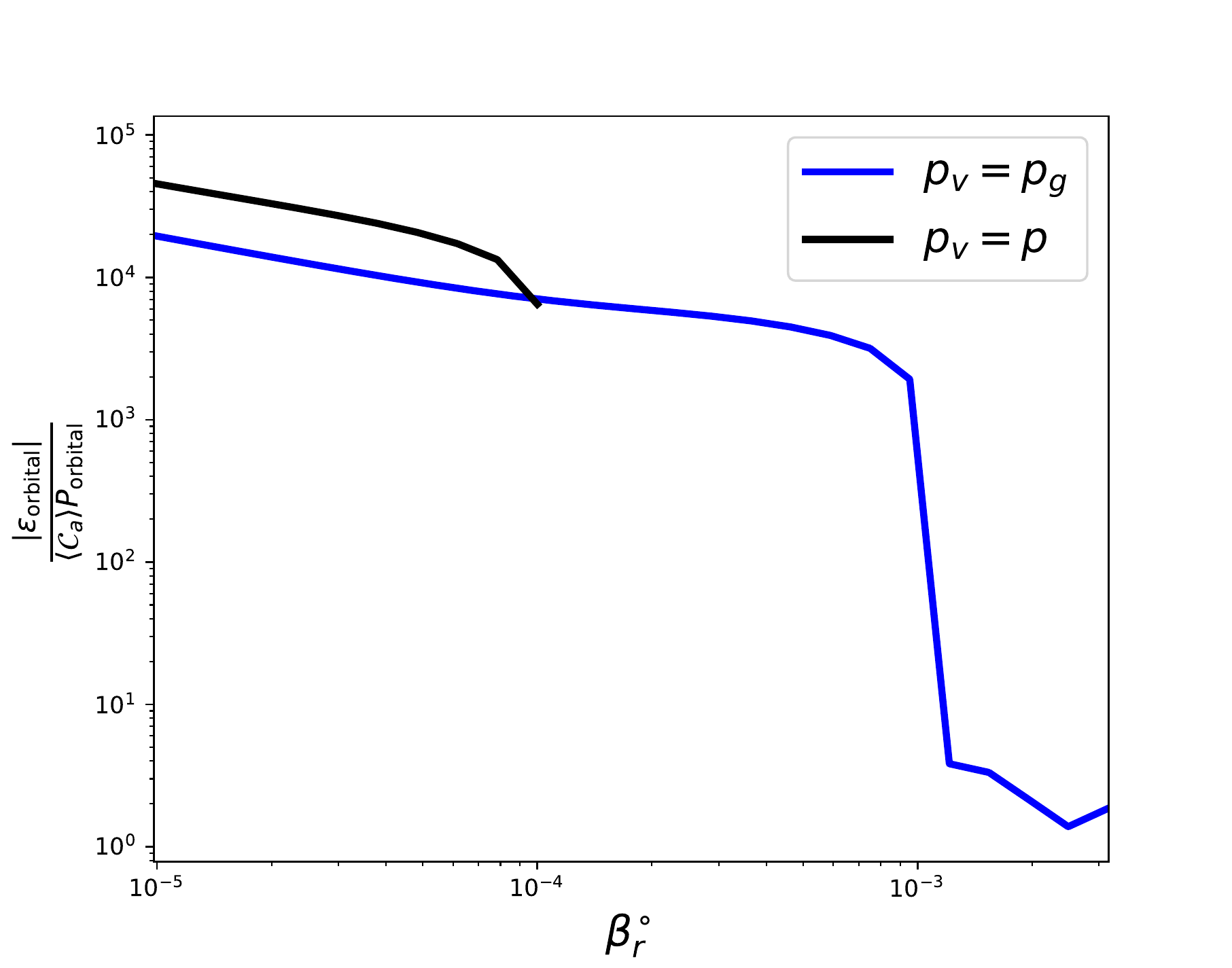}
\caption{Timescale (in orbital periods) for the orbit to radiate its entire orbital energy for the disruption of a $1 M_{\odot}$ star around a $10^{6} M_{\odot}$ black hole. For gas pressure dominated to moderately radiation pressure dominated discs the disc takes $\sim 10^4$ orbits to radiate away the orbital energy, as the discs are extremely thin. The cutoff for the $p_v=p$ line at $\beta_r^{\circ} = 10^{-4}$ is due to the onset of the unbound thermal instability. For the $p_v=p_g$ discs the thermal instability causes an abrupt drop in the time to radiate the orbital energy as the disc becomes thick.} 
\label{orbital cooling}
\end{figure}

Figure \ref{orbital cooling} shows the timescale (in orbital periods) for an orbit to radiate its entire orbital energy, given by the dimensionless quantity,

\begin{equation}
\frac{|\varepsilon_{\rm orbital}|}{\langle \mathcal{C}_a \rangle P_{\rm orbital}} = \frac{\tau^{\circ}}{16 \pi \lambda} \left(\frac{H^{\circ}}{a}\right)^{-1} \left(\frac{n a}{c} \right) \langle T^4 J^2 \rangle^{-1},
\end{equation}
where $\varepsilon_{\rm orbital} = -\frac{1}{2} M_a n^2 a^2$ is the orbital energy (per unit semimajor axis), $P_{\rm orbital}$ is the orbital period, $\langle \mathcal{C}_a \rangle$ is the orbit averaged dimensionful cooling rate (per unit semimajor axis), and $\tau^{\circ}$ and $H^{\circ}/a$ are the optical depth and aspect ratio of the reference circular disc. The reference disc has been scaled to a disruption of $1 M_{\odot}$ star around a $10^{6} M_{\odot}$ black hole. The thermally stable solutions take a large number ($\sim 10^4$) orbits to radiate their orbital energy, as the discs are very thin. For strongly radiation pressure dominated discs the disc is thermally unstable and, for $p_v=p_g$, becomes much thicker, allowing the orbital energy to be radiated on $\sim 10$ orbits. It is likely that the thermally unstable $p_v=p$ discs (not shown here) also radiate their orbital energy on much shorter timescales, potentially before they reach the unphysically large $H$ seen when we attempt to run these models to convergence.

We caution that the timescale given by $\frac{|\varepsilon_{\rm orbital}|}{\langle \mathcal{C}_a \rangle}$ should be viewed as the timescale for the orbit to change and not as a circularisation timescale. Whether the disc circularises will also depend on angular momentum transport, which is difficult to address within a fully local model. Even at fixed angular momentum a highly eccentric disc must radiate significantly more that its orbital energy to completely circularise. For instance an orbit with $e=0.9$ must radiate $\sim 4 |\varepsilon_{\rm orbital}|$ to fully circularise (at fixed angular momentum), while for an orbit with $e=0.99$ this increases to $\sim 50 |\varepsilon_{\rm orbital}|$. 

In all our models the disc radiates any energy it gained from viscous heating so cooling is fairly efficient; however the strong temperature variation around the orbit shows that this model is far from the isothermal model used in \citet{Bonnerot16} to represent an efficiently cooling disc. In some respects our highly radiation pressure dominated models are similar to the radiatively efficient polytropic model of \citet{Hayasaki16} (albeit with $\gamma=4/3$) on an orbital timescale. Although, as noted above, the disc can lose or gain entropy through radiative cooling or viscous heating  on longer timescales.

It is important to note that orbit-averaged heating equalling orbit-averaged cooling isn't an assumption of our model, it is in fact a requirement for $2\pi-$periodicity. Any imbalance would cause the disc to heat or cool until, if the disc is thermally stable, a balance between averaged heating and cooling is restored. In a real TDE, however, the orbit may change too rapidly to settle down into the periodic solutions attained here. A possible interpretation of the results of this section is that TDEs start out as a highly eccentric, very thin, disc, with orbits that evolve very slowly. This disc becomes thermally unstable, causing it to thicken, and radiates a substantial fraction of the orbital energy through a very hot, geometrically thin, region close to pericentre. This causes a rapid evolution of the disc orbits resulting in either circularisation or accretion. 

\subsection{Application to Partial Disruptions}

In this section we have mostly focused on radiation dominated discs, which are expected in full TDEs. Approximately half of tidal disruption events are expected to be partial disruptions \citep{Krolik20} and the larger disruption radius, combined with the lower disc mass, means that many of these events should be gas pressure dominated and thus exhibit the pericentre suppressed behaviour discussed in Section \ref{tde sols}. Unlike the radiation dominated solutions discussed above, these discs are both thermally stable and very thin, meaning the thin disc approximation used here is amply satisfied.

\section{Pericentre passage} \label{nozzle}

The nozzle-like structure at pericentre, involving strong heating and cooling, seen in both our pericentre-dominated models and the radiation-gas mixture, may be analogous to the `nozzle shocks' discussed in the TDE literature \citep{Evans89,Kochanek94,Guillochon14,Shiokawa15,Piran15}. These nozzle shocks are expected to arise because of the convergence of the streamlines, both horizontally and vertically, towards their common pericentre. In our model this corresponds to $q$ approaching $1$ with $E_0=0$ so that both horizontal and vertical compression occur at pericentre. In the pericentre-dominated limit the horizontal compression (while present) is less important and the nozzle-like structure is predominantly due to the vertical motion. 

\begin{figure}
\includegraphics[trim=0 0 10 10,clip,width=\linewidth]{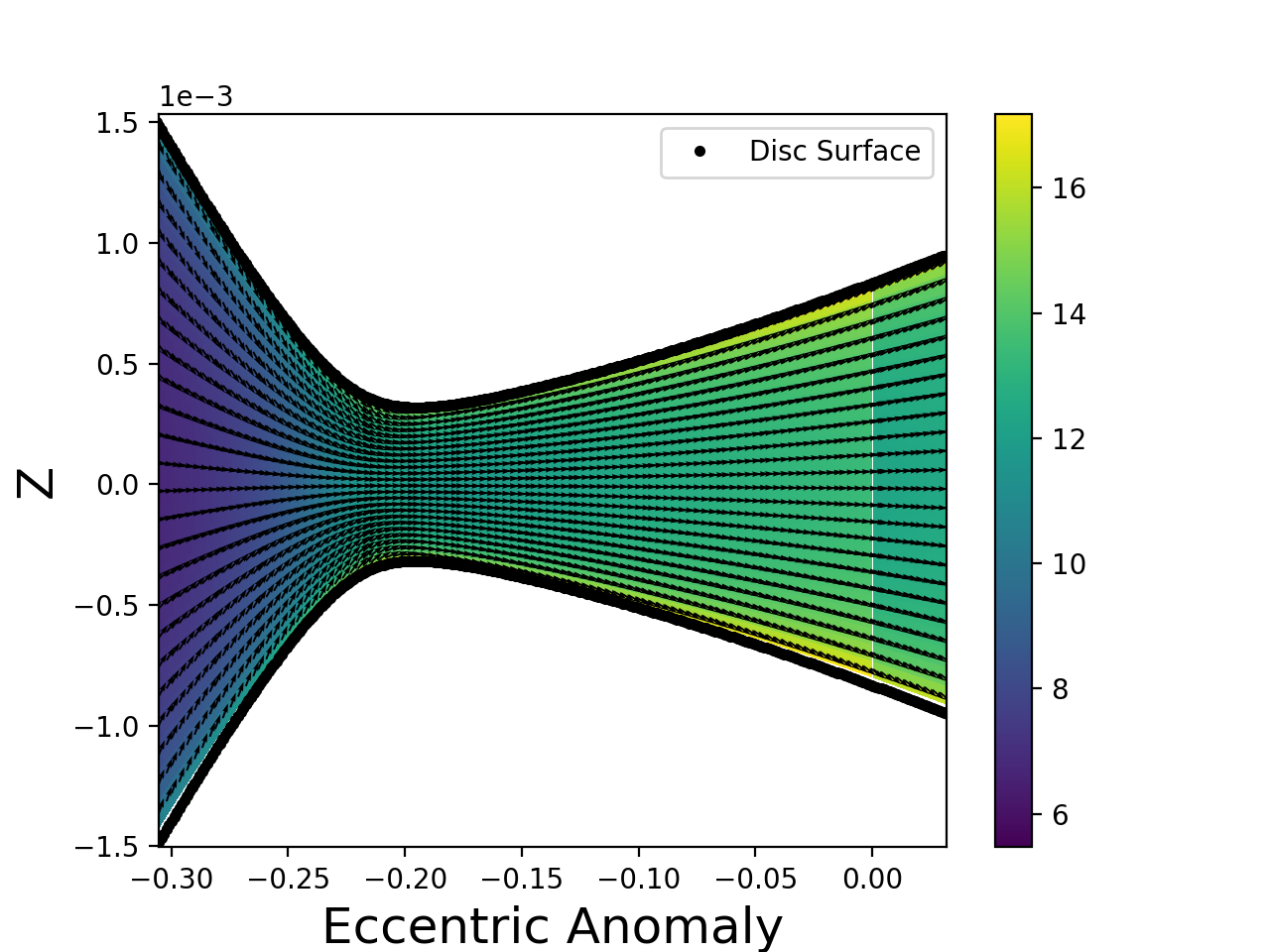}
\caption{Pericentre passage for a radiation-gas mixture with $p_v = p$, $\beta_{r}^{\circ}  = 10^{-4}$, $\alpha_s=0.01$, $\alpha_b = 0$, $e=q=0.9$,  and $E_0 = 0$. Arrows show the flow direction and colour indicates entropy. The vertical coordinate is expressed in units of $a$, using the scale height of the reference circular disc $H^{\circ}$ calculated in Appendix \ref{e q deriv}.}
\label{radmix nozzle}
\end{figure}

Figure \ref{radmix nozzle} shows the pericentre passage for a radiation-gas mixture with, $p_v = p$,  $\beta_{r}^{\circ}  = 10^{-4}$, $\alpha_s=0.01$, $\alpha_b = 0$, $e=q=0.9$,  and $E_0 = 0$. This shows a slice which follows the orbital geometry (i.e. $a=\mathrm{const}$) and is therefore different from what is typically shown in simulation papers which is typically a planar slice through pericentre.

The significant asymmetry between the disc before and after pericentre passage is caused by the dissipation of the vertical velocity in the nozzle. This feature is apparent in the abrupt change in entropy just before pericentre. Generally the more dissipative the model the more asymmetric the nozzle structure is and the earlier (relative to pericentre passage) it occurs. 

Our results differ from the expectations of a `nozzle shock'. As we include viscosity, a shock in our model would not be discontinuous and would have a finite thickness. However, the size of the (continuous) entropy jump near pericentre is dependent on $\alpha_s$ and vanishes in the inviscid limit. In fact the ideal discs possess no shock structure, strongly suggesting that the fluid can pass through the nozzle without encountering a large-scale ``nozzle shock" \footnote{This does not rule out small-scale shocks associated with turbulent motion, particularly for solutions where the viscous stress exceeds the pressure}. Our models terminate at a photosphere; it is possible there could be different dynamics above the photosphere which might include shocks, but this probably accounts for a small fraction of the mass in the disc.

Figure 11 of \citet{Shiokawa15} shows shocks just after pericentre in their hydrodynamical simulation. These shocks do not extend to the midplane, and may be present because of the finite (and large) thickness of the disc. Their Figure 11 is a snapshot taken fairly early in their simulation and may not be directly comparable to our calculations which assume the TDE has had time to settle into an eccentric disc.

\section{Discussion} \label{discuss}

\subsection{When is a $\gamma = 4/3$ perfect gas a good model of a radiation-dominated disc?} \label{rad gas perf gas comp}

In the preceding sections we have found that the $\gamma=4/3$ perfect gas model of a TDE disc behaves differently from the radiation-dominated models. In previous studies (\citet{Loeb97,Lu19,Bonnerot19}, Zanazzi \& Ogilvie 2020, submitted to MNRAS), a $\gamma=4/3$ gas has been used as a model of a radiation-dominated TDE disc. So we now consider under what circumstances this approach is valid. Generally the $\gamma=4/3$ perfect gas is used to model an adiabatic radiation pressure dominated limit so in what follows we consider a nearly adiabatic disc (i.e. $\alpha_s,\alpha_b \ll 1 $).

First consider a perfect gas with $\gamma = 4/3$ and mean molecular weight $\mu_1$. The equation of state is

\begin{equation}
p = (\mathcal{R}/\mu_1) \rho T \quad .
\label{perf eos}
\end{equation}
When this undergoes a nearly adiabatic change we have

\begin{equation}
p = K_1 \rho^{4/3} \quad ,
\label{perf adiabat}
\end{equation}
where $K_1$ is a slowly varying function. We can rearrange equations \ref{perf eos} and \ref{perf adiabat} to obtain

\begin{equation}
p = K_1^{-3} (\mathcal{R}/\mu_1)^4 T^4
\end{equation}
and

\begin{equation}
T^3 = K_1^3 (\mu_1/\mathcal{R})^3 \rho
\end{equation}

Now consider instead a mixture of a perfect gas (with mean molecular weight $\mu_2$) and black body radiation. In the radiation-dominated limit we have the equation of state

\begin{equation}
p = \frac{4 \sigma}{3c} T^4 \quad .
\label{rad eos}
\end{equation}
When this undergoes a nearly adiabatic change we have

\begin{equation}
T^3 = K_2 \rho \quad ,
\label{rad adiabat}
\end{equation}
where $K_2$ is a slowly varying function. We can rearrange equations \ref{rad eos} and \ref{rad adiabat} to obtain

\begin{equation}
p = \frac{4 \sigma}{3 c} K_2 \rho T \quad .
\end{equation}

These two systems are equivalent provided that

\begin{equation}
K_1^3 = K_2 (\mathcal{R}/\mu_1)^3
\end{equation}
and 

\begin{equation}
\frac{4 \sigma}{3 c} K_2 = (\mathcal{R}/\mu_1) \quad .
\label{molecular weight def}
\end{equation}

The first of these conditions provides a relationship between the entropies of the two systems. The second condition is more interesting. Noting the correspondence between Equations \ref{rad adiabat} and \ref{betar def}, we can rewrite Equation \ref{molecular weight def} in terms of $\beta_r$,

\begin{equation}
\frac{\mu_1}{\mu_2} = \beta_r^{-1} \quad .
\end{equation}
In the radiation-dominated limit the molecular weight of the $\gamma=4/3$ perfect gas is much lower than that of the gas in the radiation dominated disc and will correspond to some suitably weighted average of the massless photons and massive ions and electrons. As the disc becomes increasingly radiation dominated the ratio of photons to massive particles increases resulting in a decrease in the effective molecular weight of the $\gamma=4/3$ gas.

\begin{figure}
\includegraphics[trim=0 0 0 0,clip,width=\linewidth]{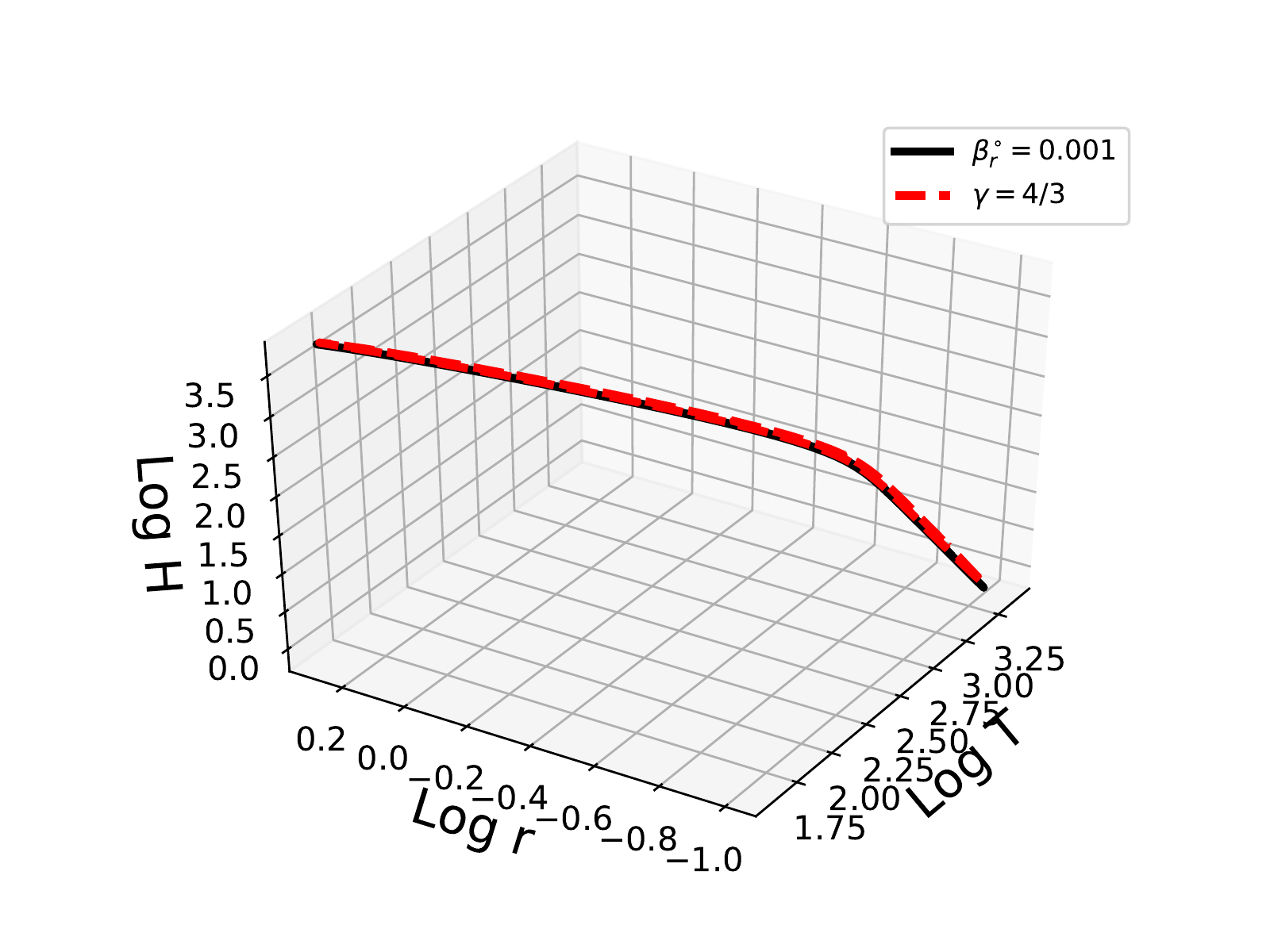}
\caption{Comparison of a $\gamma=4/3$ perfect gas model with $\alpha_s=10^{-6}$, $\alpha_b=0$ $e=q=0.9$ and $E_0 = 0$ against a radiation-gas mixture with $\beta_{r}^{\circ} = 0.001$, $\alpha_s=0.1$, $\alpha_b = 0$, $e=q=0.9$, $E_0 = 0$ and stress scaling with gas pressure. Note that this solution is radiation dominated as the resulting $\beta_r \approx 10^5$. The temperature  and scale height of the $\gamma = 4/3$ perfect gas model have been rescaled by a constant factor - this rescaling is equivalent to a change in the reference disc by selecting $K_1$ and $\mu_1$ so that the two solutions agree.} 

\label{rad gas adiabat comp}
\end{figure}

The molecular weight doesn't appear in the equations describing the dynamical vertical structure, so has no direct influence on the dynamics. Instead changing the molecular weight changes the reference state of the disc. In other words, the circular reference discs for the $\gamma=4/3$ perfect gas disc and the radiation dominated disc are different in general. For a nearly adiabatic disc where $\beta_r$ is approximately constant we can rescale the reference state so that the $\gamma=4/3$ perfect gas solution agrees with the nearly adiabatic radiation-dominated solution (see Figure \ref{rad gas adiabat comp}).

A non-adiabatic change in the radiation-dominated gas causes a change in $\beta_r$ (as the entropy $s \approx 4 \beta_r$). This corresponds to a change in the molecular weight (i.e. a compositional change) in the $\gamma=4/3$ perfect gas. In other words non-adiabatic heating and cooling in the  radiation-dominated model are equivalent to a chemical reaction in the $\gamma=4/3$ perfect gas model. This can be understood as the emission/absorption of photons by the gas resulting in a change in the ratio of photons to massive particles and, consequently, a change in the effective molecular weight $\mu_1$.

This then explains the absence of the thermal instability in the $\gamma=4/3$ perfect gas model. When the radiation-dominated disc undergoes thermal runaway the entropy (and thus $\beta_r$) increases with each orbit. In the $\gamma=4/3$ disc this would correspond to a reduction in $\mu_1$ with each orbit. However, as our equations do not allow for the evolution of $\mu_1$, the effective $\beta_r$ of the $\gamma=4/3$ disc is fixed preventing a thermal runaway. 

Most simulations use a caloric equation of state expressing internal energy $e = e (p,\rho)$ in terms of pressure and density. The thermal energy equation is then solved using the internal energy so the simulation makes no reference to either the temperature or the molecular weight, making the above a technicality. Temperatures calculated as a post-processing step using the correct equation of state (e.g. as done in \citet{Bonnerot19}, Zanazzi \& Ogilvie 2020, submitted to MNRAS) will also be correct. However, explicit cooling terms typically make reference to the temperature so the results of this section will be relevant.

\subsection{Resolution of the thermal instability}

Even in circular discs, there is a great deal of uncertainty regarding the thermal stability of radiation pressure dominated discs. Radiation pressure dominated, circular discs are thermally stable when $p_v=p_g$, but thermally unstable when $p_v=p$. Highly eccentric, radiation pressure dominated, discs behave differently and exhibit thermal instability for either stress law. This is likely due to the thermodynamics of these discs being dominated by pericentre passage. The standard scaling arguments for why circular discs should be thermally (un-)stable for a given stress law have little relevance to a highly eccentric disc. For instance, making a highly eccentric disc hotter can make it thinner, rather than thicker, at pericentre. 

There have been various suggestions about how to stabilise the thermal instability in a circular disc. The most obvious possibility is that the MRI may not be well described by an $\alpha-$disc. However recent work has demonstrated the thermal instability in radiation dominated MHD discs \citep{Latter12,Jiang13, Mishra16,Ross17}. How well this translates into an eccentric disc is highly uncertain as there have been very few studies of MRI in eccentric discs (\citet{Chan18,Dewberry20} being notable exceptions).

Another possibility that has been suggested is that the iron bound-bound transitions cause a bump in the disc opacity and can stabilise the thermal instability (e.g. \citet{Jiang16}). While this iron opacity bump is likely to be important in highly eccentric, radiation dominated, discs, its behaviour will be very different from that in a circular disc. In a circular disc the iron opacity bump sits an approximately constant height above the midplane around the entire orbit. In a highly eccentric, radiation pressure dominated, disc the iron opacity bump will be concentrated near pericentre, with a height which depends on eccentric anomaly. It is unclear, under these circumstances, what effects this will have on the disc thermodynamics; further work on this area is clearly needed. 

Finally it has been suggested by several authors that strong magnetic fields can tame the thermal instability \citep{Pessah05,Begelman07,Sadowski16a,Das18}. We have also speculated that magnetic fields may prevent the extreme compression seen in some of our models. We intend to explore these possibility in a future paper.

\subsection{Stability of the solutions} \label{stabil}

An important consideration for any steady or periodic solution of a dynamical system is its stability against perturbations. One advantage of our method is that the (in principle nonlinear) stability of our solutions to perturbations of $H, \dot{H}$ and $T$ is guaranteed as the relaxation method should only be able to find locally attracting solutions. As indicated in Section \ref{sol method} our solutions converge to the periodic solution, from a range of initial guesses, on approximately the thermal timescale. This would imply that nonlinear perturbations to our solution should damp out on the thermal timescale or faster.

For some parts of our parameter space our method fails to converge to a $2 \pi$-periodic solution. This does not necessarily mean that such a solution does not exist, but may instead indicate the $2 \pi$-periodic solution is unstable. For instance our method fails to find a $2 \pi$-periodic solution when $\beta_{r}^{\circ}$ and $\alpha_s$ are sufficiently large that the disc becomes thermally unstable. In general we find the only way to obtain highly radiation dominated discs, as expected in TDEs, is by assuming that stress scales with gas pressure. We note that a pure gas disc with $\gamma=4/3$ is thermally stable; however this is a poor model of a radiation dominated disc when dissipation and cooling are involved.

It may be that TDE discs are genuinely thermally unstable; however such an instability is unlikely to manifest itself in simulations which use simplified thermodynamics. If the thermal instability is present then the resulting thermal runaway could cause the TDE disc to become substantially thicker, potentially making it much easier to circularise the disc through enhanced dissipation. Our model is incapable of exploring this possibility, and current simulations lack the necessary thermodynamics, so a new approach would be needed to study thermally unstable TDE discs. 

Our solution method doesn't tell us about the stability of our solutions to perturbations of short wavelength (comparable or less than the scale height). So our disc structure could be unstable to such perturbations. In particular it is likely our discs are unstable to the growth of inertial waves via the parametric instability \citep{Papaloizou05a,Papaloizou05b,Wienkers18,Barker14,Pierens20}.

\subsection{Extreme behaviour during pericentre passage} \label{alpha presecription issues}

Many of our models exhibit extreme behaviour near pericentre. This is particularly true for the more radiation-dominated discs (where these are stable). In general we find that it is difficult to reverse the collapse of the fluid column during pericentre passage in the more radiation-dominated discs without encountering either extreme compression (for the nearly adiabatic discs) or unphysically strong viscous stresses (for the $\gamma = 4/3$ pure gas disc and the marginally radiation-dominated discs).

With the exception of the strongly radiation-dominated nearly adiabatic solutions, most of our solutions have a very strong vertical viscous stress at pericentre -- often comparable to or exceeding the pressure. This is particularly relevant for the pure gas models with $\gamma=4/3$, where we find that the dominant balance at pericentre is between the vertical motion and the viscous stress, which is responsible for reversing the collapse of the fluid. This issue occurs to a lesser extent in radiation-gas mixtures (with either stress model) with $\beta_r \approx 1$, where the vertical viscous stress at pericentre is typically comparable to or slightly larger than the pressure. 

Viscous stresses exceeding pressure creates a problem peculiar to turbulent/$\alpha$ stresses. This is because, if the viscous stresses are sourced from a turbulent velocity field, viscous stresses exceeding pressure imply the turbulent velocity is supersonic. Supersonic turbulence will cause shocks which should either damp the turbulent eddies or shock heat the gas causing the turbulence to become subsonic. A such the appearance of vertical viscous stresses exceeding pressure in our models should call into question the validity of the $\alpha-$prescription during pericentre passage.

For the nearly adiabatic radiation pressure dominated discs the viscous stresses are well behaved, however we encounter extreme compression during pericentre passage. This is the same as the extreme behaviour seen in the truly adiabatic discs (e.g. see \citet{Ogilvie14} and Zanazzi \& Ogilvie 2020, submitted to MNRAS). Such extreme behaviour merits caution when interpreting or accepting the results as it is likely that some neglected physical effect may regulate the behaviour near pericentre. In particular the discs we consider implicitly contain a magnetic field (responsible for the $\alpha-$viscosity) and a real magnetic field might be able to resist such an extreme compression.

In all our (stable) radiation pressure dominated discs there is thus a question as to how the magnetic field responds to the compression experienced during pericentre passage. Is the magnetic pressure capable of reversing the collapsing fluid column? Can the MRI still operate during pericentre passage, and if so how quickly can it respond to the change in pressure?

Very little work has been done on magnetic fields in highly eccentric discs. While linear growth of MRI in highly eccentric discs has been studied in \citet{Chan18} they made the assumption that the fluid motion was divergence free and thus MRI can be treated as incompressible. The presence of the breathing mode in eccentric discs causes compressional motion in the disc, so the results of \citet{Chan18} are probably only applicable to weakly eccentric discs. As is clear from our results the dominant stress at pericentre in a highly eccentric disc is that resulting from the vertical fluid motion, so the results of \citet{Chan18} are not valid in this limit. It's not clear if the turbulent stress should continue to scale with pressure, be shut  off completely (e.g. by the concentration of magnetic fields to sufficient strengths to shut off the MRI) or be in some sense ``frozen in" with the turbulence not having time to respond to the extreme conditions at pericentre.

One way of capturing the finite response timescale of MRI turbulence is through the viscoelastic model of \citet{Ogilvie00,Ogilvie01,Ogilvie02}. Using this viscoelastic prescription with a relaxation time comparable to the orbit period, the fluid should behave nearly elastically near pericentre. This will weaken the dissipative contribution to the stress that is associated with the disc turbulence. The weaker dissipation during pericentre passage may be a better model of MHD turbulence, that remains subsonic, than the simpler viscous treatment considered here.

In a future paper we will look at the behaviour of the magnetic field in a highly eccentric disc and its effects on the vertical structure. We will also look at what effect the finite response time of the MRI has on the dynamics.

\subsection{Implication of disc surface height variation for observational claims of highly eccentric TDE discs} \label{reflect imp}

One motivation for considering highly eccentric disc in TDEs is the fits to the H $\alpha$ emission made by \citet{Liu17,Cao18,Holoien19}. In these papers the authors proposed that the H $\alpha$ emission originates from the reprocessing of X-rays in a highly eccentric disc ($e\approx 0.97$ for \citet{Liu17,Cao18}; the disc of \citet{Holoien19} has more moderate eccentricity with $e \le 0.25$). The discs considered in their models are flat, being confined to a plane. As we have demonstrated in this paper this cannot be the case for a real disc as the height of the disc surface in such an eccentric disc must strongly vary around the orbit. The model of reprocessing in a flat eccentric disc would only be valid if the disc was extremely thin, such that the variation of the height of the disc surface around the orbit was negligible relative to the radial extent of the disc. This is unlikely to be the case in a real TDE disc. While present this concern is much less important for the model proposed by \citet{Hung20} owing to its much lower eccentricity.

Another overlooked factor is that from certain viewing angles the emission from pericentre will be blocked by the much thicker material at apocentre. This is probably particularly relevant to the disc model of \citet{Liu17} as their disc is both highly eccentric and close to edge on. Given the proposed orientation of the disc the much thicker material coming into pericentre should block the view to material leaving pericentre which could significantly reduce the amplitude of the redshifted peak. More work needs to be done to understand the look-angle effects on line emission from highly eccentric discs.

\section{Conclusion} \label{conc}

We have applied a non-ideal theory of eccentric discs to tidal disruption events (TDEs) discs in order to better understand their dynamical vertical structure and its variation around the elliptical orbits of a TDE disc. We have included the thermal physics appropriate to a mixture of gas and radiation undergoing viscous heating and radiative cooling. This has a significant role in determining the structure and evolution of TDE discs and will be important in understanding to what extent they circularize prior to accretion. In summary our results are:

\begin{enumerate}
\item The eccentricity in the TDE discs forces a strong variation of the scale height and entropy around the orbit.
\item The disc is out of equilibrium. It is thick where it is cold and thin where it is hot, the opposite of a disc in hydrostatic equilibrium. Additionally a highly eccentric disc can attain very high temperatures at pericentre while remaining relatively thin around the entire orbit.
\item We find a peaked structure in the entropy with a rapid change of the vertical velocity at pericentre which has much in common with the nozzle shaped flows seen in simulations. However, our results suggest the fluid can pass through the nozzle without the formation of a large-scale ``nozzle shock".
\item In many of our solutions the vertical viscous stress can exceed gas pressure. This would suggest that turbulence was supersonic -- implying a breakdown of the $\alpha-$prescription in these extreme environments.
\item Care must be taken when modelling a radiation pressure dominated disc with a perfect gas with $\gamma=4/3$, particularly when cooling or the disc temperature is involved.
\end{enumerate}

Finally and perhaps most importantly:

\begin{enumerate}
\setcounter{enumi}{5}
\item The thermal instability operates in radiation-dominated discs for both stress scaling with total pressure and stress scaling with gas pressure. In the latter case there exists an additional nearly adiabatic radiation-dominated branch which is stable and acts as an attractor for the thermal instability. It does not operate in a $\gamma=4/3$ perfect gas used as a model of a radiation-dominated disc.
\end{enumerate}

In a future paper we will explore alternative stress models which may better capture the effects of a magnetic field in an eccentric disc and lead to less extreme behaviour in the radiation dominated regime. This will also allow us to address the breakdown of the $\alpha-$prescription in highly eccentric discs. Additionally a more realistic stress model may change the thermal stability of the disc.

\section*{Acknowledgements}

We thank J. J. Zanazzi for helpful discussions and the anonymous reviewer for a number of valuable suggestions, in particular the suggestion to calculate the timescale to radiate the disc orbital energy. E. Lynch would like to thank the Science and Technology Facilities Council (STFC) for funding this work through a STFC studentship. This research was also supported by STFC through the grant ST/P000673/1. The authors thank the Yukawa Institute for Theoretical Physics at Kyoto University. Discussions during the YITP workshop YITP-T-19-07 on International Molecule-type Workshop ``Tidal Disruption Events: General Relativistic Transients" were useful to complete this work.

\section{Data availability}

The data underlying this article will be shared on reasonable request to the corresponding author.




\bibliographystyle{mnras}
\bibliography{highly_eccentric_tde} 




\appendix

\onecolumn

\section{Generalised Vertical Structure Equations} \label{general vert struct}

In this appendix we discuss the solutions to the dimensionless equations of vertical structure (Equations \ref{pressure vert strut}-\ref{eos vert strut}). These are repeated here for clarity:

\begin{equation}
\frac{d \tilde{p}}{d \tilde{z}} = - \tilde{\rho} \tilde{z} ,
\label{pressure vert strut A}
\end{equation}

\begin{equation}
\frac{d \tilde{F}}{d \tilde{z}} = \lambda \tilde{p} ,
\label{flux vert struct A}
\end{equation}

\begin{equation}
\frac{d \tilde{T}}{d \tilde{z}} = -\tilde{\rho} \tilde{T}^{ - 3} \tilde{F}_{\rm rad} ,
\label{temp vert strut A}
\end{equation}

\begin{equation}
\tilde{p} = \tilde{\rho} \tilde{T} .
\label{eos vert strut A}
\end{equation}
Here $\lambda$ is a dimensionless eigenvalue of the problem when the normalisation conditions

\begin{equation}
\int_{-\infty}^{\infty} \tilde{\rho} \, d \tilde{z} = \int_{-\infty}^{\infty} \tilde{\rho} \tilde{z}^2 \, d \tilde{z} = 1
\label{norm cond 1}
\end{equation}
are imposed. Note that these together with the vertical structure equations imply

\begin{equation}
\int_{-\infty}^{\infty} \tilde{p} \, d \tilde{z}  = 1 \quad .
\label{norm cond 2}
\end{equation}

At the upper surface of the disc the density and pressure are assumed to vanish. Defining $\tilde{z}_s$ to be the location of the upper surface, with $\tilde{p} (\tilde{z}_s) = \tilde{\rho} (\tilde{z}_s) = 0$, we have

\begin{equation}
2 \tilde{F} (\tilde{z}_s) = \lambda \quad .
\end{equation}
Thus $\lambda$ is a dimensionless cooling rate. 

The total dimensionless heat flux is $\tilde{F}=\tilde{F}_{\rm rad}+\tilde{F}_{\rm ext}$. We assume that heat is only transported radiatively in the disc photosphere so that $\tilde{F}_{\rm ext} (z_{s}) = 0$ and $\tilde{F} (\tilde{z}_s) = \tilde{F}_{\rm rad} (\tilde{z}_s)$. In the pure gas case we take $\tilde{F} = \tilde{F}_{\rm rad}$ throughout and Equations \ref{pressure vert strut A}-\ref{eos vert strut A} can be solved numerically to determine that $\lambda = 3.213$. 

For a radiation+gas mixture there is an additional contribution to the heat flux $\tilde{F}_{\rm ext}$ due to vertical mixing, either from convection \citep{Bisnovatyi76}, or from the disc turbulence. This mixing is assumed to set up a vertically isentropic structure, which for a radiation dominated disc, corresponds to a polytrope with index $3$ (this is a similar model to that of \citet{Ferreira08} who looked at eccentricity propagation in a black hole disc). For a radiatively cooled, gas-dominated disc the vertical structure is close to the polytrope with index $3$. As such a polytropic vertical structure with index $3$ should be a good approximation to the radiation+gas mixture case for a constant opacity disc. This model has $\tilde T^3/\tilde\rho=\text{const}$ and therefore $\beta_r$ is independent of $z$, which is required to maintain separability. The pressure and density for the polytrope are

\begin{equation}
\tilde{p} = \tilde{p}_0 \left[1 - \left(\frac{\tilde{z}}{\tilde{z}_s} \right)^2\right]^4 , \quad \tilde{\rho} = \tilde{\rho}_0 \left[1 - \left(\frac{\tilde{z}}{\tilde{z}_s} \right)^2\right]^3 \quad .
\end{equation}
The normalisation conditions (Equations \ref{norm cond 1}-\ref{norm cond 2}) set $\tilde{z}_s=3$, $\tilde{p}_{0}=\frac{105}{256}$, $\tilde{\rho}_0=\frac{35}{96}$. The flux within the disc can be obtained by directly integrating Equation \ref{flux vert struct},

\begin{equation}
\tilde{F} = \lambda \tilde{p}_0 \tilde{z} \left[1 - \frac{4}{3} \left(\frac{\tilde{z}}{\tilde{z}_s} \right)^2 + \frac{6}{5} \left(\frac{\tilde{z}}{\tilde{z}_s} \right)^4 - \frac{4}{7} \left(\frac{\tilde{z}}{\tilde{z}_s} \right)^6 + \frac{1}{9} \left(\frac{\tilde{z}}{\tilde{z}_s} \right)^8 \right] \quad ,
\label{flux explicit}
\end{equation}
where we have made use of the symmetry condition $\tilde{F} (0) = 0$. 

The boundary condition at the upper disc surface $\tilde{F} (\tilde{z}_s) = \tilde{F}_{\rm rad} (\tilde{z}_s)$ can be satisfied by evaluating Equation \ref{flux explicit} at $\tilde{z} = \tilde{z}_s$ and substituting into Equation \ref{temp vert strut}. The boundary condition is satisfied provided that

\begin{equation}
\lambda = \frac{315}{64} \frac{\tilde{p}^3_0}{\tilde{\rho}_0^5 \tilde{z}_s^2} = \frac{6561}{1120} \quad .
\end{equation}
This is higher than the pure gas as convective/turbulent mixing allows for more efficient heat transport near the disc midplane.


\section{Heating function} \label{heat func expl}

In this appendix we derive the dimensionless heating rate $f_{\mathcal{H}}$ for an eccentric $\alpha-$disc. We will find that $f_{\mathcal{H}}$ can be written as a horizontal part depending only on the orbital geometry (an explicit function of $E$ with coefficients involving $\alpha_{s,b}$) together with

\begin{equation}
2 \alpha_s \left( \frac{\dot{H}}{H} \right)^2 + \left(\alpha_b - \frac{2}{3} \alpha_s \right) \left(\frac{\dot{J}}{J} + \frac{\dot{H}}{H} \right)^2 \quad .
\end{equation}

The viscous stress tensor is given by

\begin{equation}
T^{a b} = 2 \mu_s S^{a b} + (\mu_b - \frac{2}{3} \mu_s) \nabla_c u^c g^{a b} \quad ,
\end{equation}
where the rate-of-strain tensor is 

\begin{equation}
S^{a b} = \frac{1}{2} \left(\nabla^a u^b + \nabla^b u^a \right)
\end{equation}
and $g^{a b}$ is the metric tensor.

The viscous heating rate per unit volume is given by

\begin{equation}
\mathcal{H} = T^{a b} S_{a b} = 2 \mu_s S^{a b} S_{a b} + \left(\mu_b - \frac{2}{3} \mu_s \right) S^2
\end{equation}
where $S=S_a^a = \nabla_a u^a = \Delta + \frac{\dot H}{H}$ is the velocity divergence. Assuming an $\alpha$ viscosity (with $\omega_{\rm orb} = n$) and using the relationship  $\mathcal{H} = f_{\mathcal{H}} n p_v$ we obtain for the dimensionless heating rate

\begin{equation}
f_{\mathcal{H}} = 2 \alpha  s^{a b} s_{a b} + \left(\alpha_b - \frac{2}{3} \alpha \right) \left(\frac{\dot{J}}{J} + \frac{\dot{H}}{H} \right)^2
\end{equation}
where we have introduced a dimensionless rate-of-strain tensor such that $S^{a b}= n s^{ab}$. It now remains to determine $s^{a b} s_{a b}$. This transforms as a scalar under change of variables so it is straightforward to take the expressions for the components of $s^{a b}$ and $s_{a b}$ in the $(\lambda, \phi)$ coordinate system from the appendix of \citet{Ogilvie01} and obtain 

\begin{align}
\begin{split}
s^{a b} s_{a b} &= \frac{(1 + e c)^2 }{8 \left(1 - e^2\right)^3 [1 + (e-\lambda e_{\lambda}) c -e \lambda \varpi_{\lambda} s]^2} \Biggl\{c^4 e^2 \left[e^2 \left(8 \lambda \varpi_{\lambda}^2+1\right)-4 e \lambda e_{\lambda}+4 \lambda e_{\lambda}^2\right] + 8 c^3 e \left[e^2 \left(2 \lambda \varpi_{\lambda}^2+1\right)-3 e \lambda e_{\lambda}+2 \lambda e_{\lambda}^2\right] \\ 
&+2 c^2 \left[e^4 \left(8 \lambda \varpi_{\lambda}^2+1\right) s^2-4 e^3 s (\lambda e_{\lambda} s+3 \lambda \varpi_{\lambda})+e^2 \left(4 \lambda e_{\lambda}^2 s^2+8 \lambda e_{\lambda} \lambda \varpi_{\lambda} s+4 \lambda \varpi_{\lambda}^2+11\right)-22 e \lambda e_{\lambda}+8 \lambda e_{\lambda}^2\right] \\
&+ 8 c \left[4 e^3 \lambda \varpi_{\lambda}^2 s^2-6 e^2 \lambda \varpi_{\lambda} s+e (2 \lambda e_{\lambda} \lambda \varpi_{\lambda} s+3)-3 \lambda e_{\lambda}\right] + e^4 \left(8 \lambda \varpi_{\lambda}^2+1\right) s^4-4 e^3 \lambda e_{\lambda} s^4 \\
 & +2 e^2 s^2 \left(2 \lambda e_{\lambda}^2 s^2-8 \lambda e_{\lambda} \lambda \varpi_{\lambda} s+8 \lambda \varpi_{\lambda}^2-1\right)+4 e s (\lambda e_{\lambda} s-6 \lambda \varpi_{\lambda})+8 \lambda e_{\lambda}^2 s^2+9 \Biggr\} + \left(\frac{\dot{H}}{H} \right)^2
\end{split}
\label{lam phi shear heat}
\end{align}
where we have defined $c := \cos f$, $s := \sin f$ and the subscript $\lambda$ denotes a derivative with respect to $\lambda$.

$s^{a b} s_{a b}$ in the $(a,E)$ coordinate system can be obtained by substituting the appropriate expressions for $\lambda$, $\phi$, $\lambda e_{\lambda}$, $\lambda \varpi_{\lambda}$ using a symbolic algebra package such as Mathematica. The resulting expression is quite a bit longer than Equation \ref{lam phi shear heat} so we don't include it here.
 
\section{Typical orbital elements and reference disc for a TDE} \label{e q deriv}

In this appendix we calculate the typical orbital elements and $H^{\circ}$ for a TDE disc shortly after disruption. To do this fully self-consistently we should solve for $e$ and $\varpi$ as a function of $a$ using the eccentric disc theory of \citet{Ogilvie01} (adapted for our coordinate system). This has been done recently by Zanazzi \& Ogilvie (2020, submitted to MNRAS) but is beyond the scope of this paper. Instead we consider the fluid orbits immediately after disruption.
 
Consider a star on a parabolic orbit which is instantaneously disrupted at pericentre. The fluid is assumed to exit pericentre with a uniform velocity (equal to the star's velocity at the moment of disruption) and the effects of self gravity and pressure are neglected. As the star is radially extended, the fluid is emplaced onto a range of bound and unbound orbits. The centre of the star (which continues along the original stellar orbit) follows a parabolic reference orbit with pericentre distance $R_p$ and velocity $v_p$ related by
 
\begin{equation}
\frac{1}{2} v_p^2 - \frac{G M_{\bullet}}{R_p} = 0 \quad .
\end{equation}
 
Fluid elements all have the same pericentre velocity $v_p$ but have a range of pericentre distances $R$ given by
 
\begin{equation}
R = R_p (1 + x) \quad .
\end{equation}
 
Their specific energy is
 
\begin{equation}
\mathcal{E} = \frac{v_p^2}{2} - \frac{G M_{\bullet}}{R} = -\frac{G M_{\bullet}}{2 a}
\end{equation}
which we can use to obtain an equation for the semimajor axis and eccentricity parameterised in terms of $x$,
 
\begin{equation}
a = -\frac{R_p}{2} x^{-1} (1 + x) \quad ,
\label{semimajor axis dist}
\end{equation}
 
\begin{equation}
e = 1 - \frac{R}{a} = 1 + 2 x \quad .
\end{equation}
From these we can calculate $q$,
 
\begin{equation}
q = \frac{1}{1 - 2 x} \approx 1 + 2 x + O(x^2)
\end{equation}
and so we conclude that for the early TDE disc we have
 
\begin{equation}
e \approx q
\end{equation}
and 
 
\begin{equation}
e = 1 - \frac{2 R_p}{2 a + R_p} \approx 1 - R_p/a \quad.
\label{ecc profile}
\end{equation}
 
For the unbound debris with $x>0$ we have $q>1$. Normally this would indicate an orbital intersection, however the $(a,E)$ coordinate system is not well defined for unbound orbits. Using the $(\lambda,\phi)$ coordinate system, or simply by plotting the orbits, it's apparent that the unbound debris has no orbital intersection for the eccentricity profile considered.
 
This eccentricity profile is for the orbits immediately after disruption. However, our model requires the fluid to be spread over the range in eccentric anomaly in order for it to form a coherent disc. This spread in $E$ is initially prevented by the stream self gravity \citep{Kochanek94,Guillochon14} so some degree of circularisation must have occurred. This will modify the eccentricity profile away from Equation \ref{ecc profile}. In the limit where apsidal precession is weak this circularisation will be weak and its main effect will be in breaking the hold self gravity has on the stream thickness and allowing the stream to spread around the orbit. In most of our models we have chosen a slightly lower eccentricity ($e=0.9$) than the eccentricity of the most bound debris for the disruption of a solar like star calculated with equation \ref{ecc profile} ($e=0.98$).
 
\begin{figure}
\includegraphics[trim=0 0 0 0,clip,width=\linewidth]{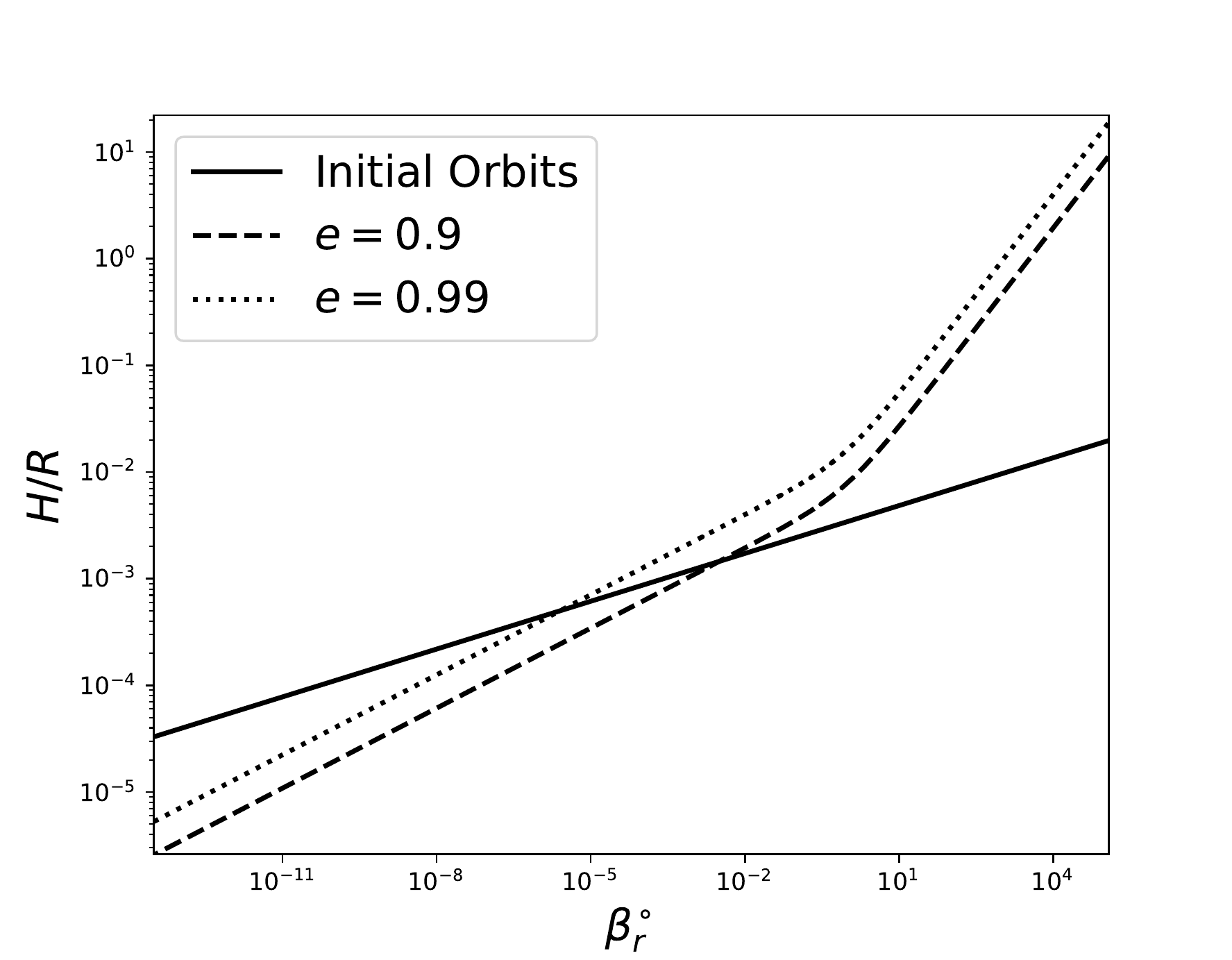}
\caption{Reference $H/R$ against reference $\beta_r$ from the disruption of a $1 M_{\odot}$ star around a $10^6 M_{\odot}$ black hole at the tidal radius. The solid line is assuming the debris remains on the post disruption orbits given by Equation \ref{ecc profile}. The dashed and dotted lines keep $e$ and $q$ fixed as $\beta_r^{\circ}$ is varied. Stress is assumed to scale with gas pressure ($p_v = p_g$).}
\label{circ plot}
\end{figure}

For a given black hole mass and $R_p$, Equation \ref{semimajor axis dist} allows us to calculate the mean motion $n$ of the disc orbits. This is required to determine the scale height of the reference circular disc (important for determining if the disc is thin). We can rearrange the equations for hydrostatic and thermal balance, along with the equation of state, to obtain an equation for $H^{\circ}$ in terms of $\beta_r^{\circ}$,

\begin{equation}
H^{\circ} = \left[ \frac{4 \lambda}{9 \alpha_s} \frac{16 \sigma}{3 \kappa n^7} \left(\frac{\mathcal{R}}{\mu}\right)^{4} \left( \frac{3 c}{4 \sigma} \right)^{2} (\beta_{r}^{\circ})^2 (1 + \beta_r^{\circ})^3 \right]^{1/8} \quad .
\end{equation}
Figure \ref{circ plot} shows how $H^{\circ}/a$ varies with $\beta_r^{\circ}$ for typical TDE streamlines.

\bsp	
\label{lastpage}
\end{document}